\documentclass[
superscriptaddress,
onecolumn,
preprint,
 amsmath,
 amssymb,
 aps,
 prb,
]{revtex4-2}

\usepackage{graphicx}
\usepackage{dcolumn}
\usepackage{bm}
\usepackage{siunitx}
\usepackage{hyperref}

\begin{document}

\title{Slip and friction at fluid-solid interfaces: Concept of adsorption layer}

\author{Haodong Zhang}
\author{Fei Wang}%
\email{fei.wang@kit.edu}
\affiliation{%
 Institute of Applied Materials-Microstructure Modelling and Simulation, 
Karlsruhe Institute of Technology (KIT), Strasse am Forum 7, Karlsruhe 76131, Germany.}
\affiliation{Institute of Nanotechnology, Karlsruhe Institute of Technology (KIT), 
Hermann-von-Helmholtz-Platz 1, Eggenstein-Leopoldshafen 76344, Germany}%
\author{Britta Nestler}
\affiliation{%
 Institute of Applied Materials-Microstructure Modelling and Simulation, 
Karlsruhe Institute of Technology (KIT), Strasse am Forum 7, Karlsruhe 76131, Germany.}
\affiliation{Institute of Nanotechnology, Karlsruhe Institute of Technology (KIT), 
Hermann-von-Helmholtz-Platz 1, Eggenstein-Leopoldshafen 76344, Germany}%
\affiliation{
Institute of Digital Materials Science, Karlsruhe University of Applied Sciences, 
Moltkestrasse 30, Karlsruhe 76133, Germany}%

\date{\today}

\begin{abstract}
When a fluid flows past a solid surface, 
its macroscopic motion arises from a subtle interplay 
between microscopic hydrodynamic and thermodynamic effects 
at the fluid–solid interface.
Classical hydrodynamic models often rely on an unphysical no-slip boundary condition or an arbitrarily prescribed slip length $l_{\scriptstyle \text{s}}$, 
yet both approaches lack a rigorous physical foundation.
This work introduces the concept of an Adsorption Layer (AL), an interfacial region of thickness $\delta l$, where fluid-solid molecular interactions regulate both surface adsorption/depletion and interfacial slip.
By applying the energy minimization principle, we derive balance equations within the AL that couple fluid-solid friction, viscous stresses, and surface adsorption dynamics. 
This framework establishes a self-consistent thermodynamic coupling between the AL and the bulk fluid, unlike conventional sharp-interface models. 
A key finding is the often-overlooked role and coupling of pressure and chemical potential gradients in the direction normal to the interface. 
This theoretical advance successfully explains the confinement-induced enhancement of water slippage in carbon nanotubes, quantitatively agreeing with molecular dynamics and experimental data—an effect classical slip models fail to reproduce. Furthermore, when extended to binary liquids, the theory captures spatial variations in slip velocity near moving contact lines, highlighting the role of interfacial friction in shaping local flow.
Our results demonstrate that the slip length is not a fixed material constant but rather an emergent, geometry- and composition-dependent property arising from coupled interfacial thermodynamics and hydrodynamics. This framework provides a physically grounded description of interfacial momentum transfer, with significant implications for microfluidics and surface engineering.
\end{abstract}

\maketitle

\section{Introduction}
\begin{figure}
\includegraphics[width=0.9\textwidth]{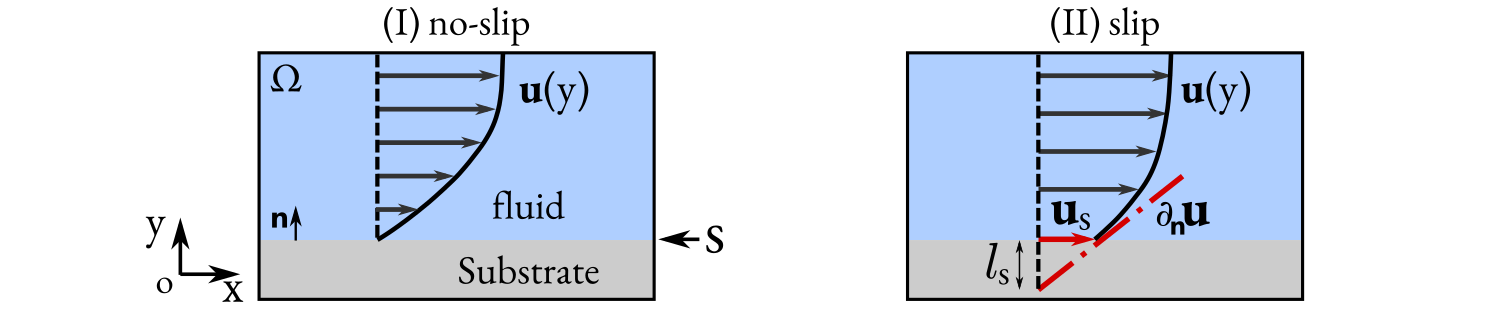}
\centering
\caption{\textcolor{black}{(I) Fluid velocity distribution $\mathbf{u}(y)$ with no-slip boundary condition at fluid-substrate contacting surface S.
(II) Slip boundary condition enables non-zero fluid velocity $\mathbf{u}_{\scriptstyle \text{s}}$ at S.
The Navier slip length is mathematically defined as $l_{\scriptstyle \text{s}}=\mathbf{u}_{\scriptstyle \text{s}}/\partial_{\text{n}} \mathbf{u}$.}
}
\label{fig:slip}
\end{figure}

When a moving fluid interacts with a solid substrate, 
the velocity field near the surface forms a boundary layer, 
which can be broadly categorized into the following classical regimes.

\emph{(I) No-Slip Boundary Condition}:
From a macroscopic viewpoint, 
extensive experimental and theoretical investigations~\citep{lauga2005microfluidics,luchini2013linearized,wen2022steady} 
support the no-slip condition $\textbf{u}_{\scriptstyle \text{s}} = \textbf{0}$,
implying zero fluid velocity at the fluid–solid interface S, 
as illustrated in figure~\ref{fig:slip}(I). 
This condition has long been regarded as a cornerstone of continuum fluid mechanics. 
With the emergence of computational fluid dynamics, 
it became a standard modeling assumption due to its strong predictive accuracy in a wide range of flow problems~\citep{qian2008scaling,yue2012phase,yue2020thermodynamically,ding2023heat}.
However, this assumption introduces several conceptual inconsistencies.
First, the solid–fluid boundary under the no-slip condition does not dissipate mechanical energy. 
Taking laminar Poiseuille flow as an example, 
the total energy dissipation rate can be expressed as
\begin{align*}
   \frac{\mathrm{d}\mathcal{L}}{\mathrm{dt}} 
= -\int_{\scriptstyle \Omega} \eta_{\scriptscriptstyle\,}\nabla\mathbf{u}:\!\nabla\mathbf{u}\,\mathrm{d}\Omega\leq 0,
\end{align*}
indicating the energy is dissipated solely 
within the bulk fluid domain $\Omega$ 
through viscous stresses governed 
by the dynamic viscosity $\eta$
and the velocity gradient $\nabla\mathbf{u}$.
Second, numerous nanoscale experiments have revealed 
measurable slip velocity at the fluid-solid interface~\citep{neto2005boundary,myers2011slip,maali2012measurement,xie2018fast},
a finding further corroborated by molecular dynamics (MD) simulations~\citep{qian2003molecular,qian2006variational,voronov2006boundary,kumar2012slip,ramos2016hydrodynamic,alipour2019molecular}. 
These results challenge the universality of the no-slip condition 
and underscore the need for refined boundary models.\vspace{0.5em}

\textcolor{black}{\emph{(II) Navier Slip Boundary Condition (NBC)}: This concept of interfacial slip dates back to 
the seminal works of~\cite{navier1822memoire,maxwell1879vii}.
Here slip is characterized 
by the slip length $l_{\scriptstyle\text{s}}$~\citep{lauga2003effective,kusumaatmaja2016moving,yariv2023effective}, defined by 
\begin{align}
    \textbf{u}_{\scriptstyle \text{s}} = l_{\scriptstyle\text{s}}\,\partial_{\scriptstyle \text{n}} \textbf{u},~\label{eq:NBC}
\end{align}
where $\partial_{\scriptstyle\text{n}}\textbf{u}$ denotes the velocity gradient normal to the solid substrate;
see figure~\ref{fig:slip}(II).
Experimental observations show that $l_{\scriptstyle\text{s}}$ can vary from nanometers to micrometers~\citep{joseph2005direct,muralidhar2011influence},
depending on interfacial properties and topography.
In confined geometries, such as carbon nanotubes, 
reported slip lengths can be several orders of magnitude larger than those observed in macroscopic systems~\citep{secchi2016massive}.
Under the slip boundary condition, the total energy dissipation rate becomes
\begin{align}
  \frac{\mathrm{d}\mathcal{L}}{\mathrm{dt}}  
=-\int_{\scriptstyle \text{S}} (\eta/l_{\scriptstyle \text{s}})_{\scriptstyle \,} \mathbf{u}_{\scriptstyle \text{s}}^{\scriptstyle 2} 
\,\mathrm{dS}
 -\int_{\scriptstyle \Omega} \eta_{\scriptscriptstyle\,}\nabla\mathbf{u}:\!\nabla\mathbf{u}\,\mathrm{d}\Omega\leq 0.\label{eq:dis_slip}
\end{align}
Here, both bulk and interfacial dissipation contribute to the overall energy loss.
As $l_{\scriptstyle \text{s}}$ increases, interfacial dissipation weakens accordingly.
In practice, the slip length $l_{\scriptstyle \text{s}} $ is not known \textit{a priori} 
and must be determined indirectly—through flow-rate measurement, interfacial force experiment, or MD simulation.
To connect interfacial friction to hydrodynamic quantities, $l_{\scriptstyle \text{s}}$ is often related to 
the intrinsic friction coefficient 
$\lambda_{\scriptscriptstyle\text{intr}}$ via
\begin{align}
    l_{\scriptstyle \text{s}} 
    =\frac{\;\eta\;}{\lambda_{\scriptscriptstyle\text{intr}}}.\label{eq:L_s}
\end{align}
Accordingly, the first term in \eqref{eq:dis_slip} represents the dissipation by the interfacial friction force,
\begin{equation*}
    \textbf{F} = -\lambda_{\scriptscriptstyle\text{intr}}\mathbf{u}_{\scriptstyle \text{s}},
\end{equation*}
where $\lambda$ can be evaluated from equilibrium MD simulations 
using the Green–Kubo relation~\citep{bocquet1994hydrodynamic},
\begin{align}
    \lambda_{\scriptscriptstyle\text{intr}} = \frac{1}{\text{S}k_{\scriptscriptstyle \!B}T}\!\int_{\scriptstyle 0}^{\scriptstyle \infty}\!\!\!\mathrm{dt} \,\langle \mathbf{F}(\mathbf{x},\text{t}), \,\mathbf{F}(\mathbf{x},0)\rangle,\label{eq:lambda_gk}
\end{align}
in which $k_{\scriptscriptstyle \!B}$ and $T$
denote the Boltzmann constant and temperature, respectively.
This method has been successfully applied to quantify slippage 
in diverse interfacial systems, 
from hydrophobic surfaces to carbon nanotubes~\citep{secchi2016massive,cui2023enhanced}.
Nonetheless, the computational cost of MD simulations 
shows their applicability only to nanoscale or submicron systems; 
they also remain impractical for multicomponent flows at larger scales.
With the friction coefficient, 
NBC~\eqref{eq:NBC} is also expressed as
\begin{align*}
    \lambda_{\scriptscriptstyle\text{intr}}\mathbf{u}_{\scriptstyle \text{s}} 
    =\eta\nabla \mathbf{u}\cdot\mathbf{n}.
\end{align*}}

\textcolor{black}{(III) \emph{Generalized Navier boundary condition (GNBC)}: 
A complementary mean-field perspective treats the fluid–solid interface 
as a sharp boundary where intermolecular interactions induce depletion/adsorption layers~\citep{de1981polymer,wang2023thermodynamically,zhang2024wetting},
thereby modulating wetting and dewetting dynamics at the macroscopic level.
During the wetting induced flow, these interactions also 
generate hydrodynamic damping near the wall, 
producing velocity gradients consistent with interfacial slip~\citep{xia2024unveiling,hadjiconstantinou2024molecular}.}

\textcolor{black}{Building on molecular insights, 
\cite{qian2003molecular}
derived GNBC for a two-dimensional coordinate system (x,n), 
relating the slip velocity $\mathbf{u}_{\scriptstyle \text{s}}=(\text{u}_{\scriptstyle \text{s}},\text{v}_{\scriptstyle \text{s}})$ to fluid-solid interfacial force balance as,
\begin{align}
    \lambda_{\scriptscriptstyle\text{intr}} \text{u}_{\scriptstyle \text{s}} = -\eta\,(\partial_{\scriptstyle \text{n}\,} \text{u}_{\scriptstyle \text{s}} + \partial_{\scriptstyle \text{x}} \text{v}_{\scriptstyle \text{s}}) + \tilde{\mu}_{\scriptstyle \text{s}\,}\partial_{\scriptstyle \text{x}} c_{\scriptstyle \text{s}},\label{eq:slip_qian}
\end{align}
where the non-uniform surface composition $c_{\scriptstyle \text{s}}$ leads to 
an extra capillary term into NBC.
Here, the surface chemical potential of sharp interface model is defined as $\tilde{\mu}_{\text{s}}=\kappa\nabla c_{\text{s}}\cdot \mathbf{n} + \partial \gamma/\partial c_{\text{s}}$
with $\gamma$ denoting the fluid-solid interfacial tension.
The surface composition then evolves by the dynamic wetting boundary condition, following the mean curvature flow as
\begin{align}
    \frac{\partial c_{\scriptstyle \text{s}}}{\partial\text{t}}=-\tau \tilde{\mu}_{\text{s}},~\label{eq:conc_qian}
\end{align}
where the kinetic parameter $\tau$ controls the surface diffusion.
Including the composition diffusive dissipation, 
the total energy dissipation rate becomes,
\begin{align*}
 \frac{\mathrm{d}\mathcal{L}}{\mathrm{dt}} 
=-\int_{\scriptstyle \text{S}} \lambda_{\scriptscriptstyle\text{intr}}
\text{u}_{\scriptstyle \text{s}}^{\scriptstyle 2} 
\,\mathrm{dS}
-\int_{\scriptstyle \text{S}} \tau_{\scriptscriptstyle \,} \tilde{\mu}_{\scriptstyle \text{s}}^{\scriptstyle 2}
\,\mathrm{dS}
 -\int_{\scriptstyle \Omega} \eta_{\scriptscriptstyle\,}\nabla\mathbf{u}:\!\nabla\mathbf{u}\,\mathrm{d}\Omega
 -\int_{\scriptstyle \Omega} M_{\scriptscriptstyle\,}(\nabla\mu)^2\,\mathrm{d}\Omega\leq 0.
\end{align*}
The GNBC model has been validated against MD simulations of binary Couette flows, 
showing quantitative agreement.
Nevertheless, a key limitation of Qian’s formulation is that
interfacial energy dissipation is established under steady force balance conditions,
implying that~\eqref{eq:slip_qian} 
represents a steady-state slip law rather than a fully dynamic one. }\vspace{0.5em}

\textcolor{black}{(IV) \emph{Dynamic slip boundary condition}: The slippage under non-equilibrium condition 
was first systematically analyzed in the seminal work of~\cite{bedeaux1986nonequilibrium}.
By evaluating the total surface energy dissipation 
at the fluid–solid dividing surface, 
he derived the interfacial momentum balance as,
\begin{gather}
\rho_{\scriptstyle \text{s}}\frac{\mathrm{d}\mathbf{u}_{\scriptstyle \text{s}}}{\mathrm{dt}}
= \nabla_{\!\scriptstyle\text{s}} \cdot \big[\gamma(\mathbf{I}-\mathbf{n}\otimes\mathbf{n})+\eta\,(\nabla_{\!\scriptstyle\text{s}} \mathbf{u}_{\scriptstyle \text{s}} + \nabla_{\!\scriptstyle\text{s}} \mathbf{u}_{\scriptstyle \text{s}}^{\scriptscriptstyle\text{T}})\big]
+\mathbf{F}.~\label{eq:slip_bedeuax}
\end{gather}
where $\gamma$ denotes the surface tension
and \textbf{F} represents the net external force acting along the interface.
The surface gradient operator is defined as
$\nabla_{\!\scriptstyle\text{s}}:=\nabla-\textbf{n}(\textbf{n}\cdot\nabla)$.
Here, the surface density follows the material conservation law~\citep{bedeaux1986nonequilibrium},
\begin{align}
\partial \rho_{\scriptstyle \text{s}}/\partial\text{t}
+\nabla_{\!\scriptstyle\text{s}}\cdot(\rho_{\scriptstyle\text{s}\,}\mathbf{u}_{\scriptstyle\text{s}})=0,~\label{eq:density_bedeaux}
\end{align}
which is more rigorous than the non-conserved~\eqref{eq:conc_qian}.
Using the entropy production principle,~\cite{bothe2016interface} subsequently obtained the same result from 
the second law of thermodynamics, 
employing a linear closure approach to formulate 
a more generalized interfacial velocity boundary condition
(see also~\cite{bothe2022sharp} for detailed derivations).
At the steady state $\partial\mathbf{u}_{\scriptstyle \text{s}}/\partial\text{t}=0$ 
and neglecting the inertia term $\mathbf{u}_{\scriptstyle \text{s}}\cdot\nabla\mathbf{u}_{\scriptstyle \text{s}}$, 
the momentum balance~\eqref{eq:slip_bedeuax} reduces to GNBC~\eqref{eq:slip_qian}.}

\textcolor{black}{In these foundational works~\citep{bedeaux1986nonequilibrium,qian2003molecular,bothe2016interface}, 
the emphasis lies primarily on deriving the governing interfacial equations.
Determining the actual slippage at the fluid–solid interface, 
however, often requires numerical simulations.}

\textcolor{black}{To complement simulation-based approaches, 
earlier analytical efforts focused on the slip at fluid-solid interface always adopt
either of the above four formulations.
In the seminal wedge flow model by~\cite{moffatt1964viscous,huh1971hydrodynamic}, 
analytical solutions for the local flow field
of the liquid-air-solid system 
near the contact line is obtained under the no-slip boundary condition.
However, this formulation produces a shear-stress singularity at the contact line.}

\textcolor{black}{
To resolve this issue,~\cite{hocking1977moving} incorporated 
the NBC at the liquid–solid interface, 
obtaining a finite frictional force associated with the motion of the fluid 
along the solid substrate.
Based on this framework, subsequent studies~\citep{snoeijer2006free, chan2020cox,kansal2025viscoelastic} 
developed the generalized lubrication theory, 
extending Hocking’s results 
to the slippage of both Newtonian and viscoelastic fluids.
}

\textcolor{black}{In addition, 
all these analytical models treat the fluid as a compositionally homogeneous medium, 
neglecting interfacial variations in density or concentration.
To incorporate surface depletion/adsorption effects into the wedge-flow framework,~\cite{shikhmurzaev1993moving} 
proposed the interface formation model, 
which corresponds to the steady state of dynamic slip boundary condition
to include surface density variations.
The steady state of~\eqref{eq:slip_bedeuax} is expressed in two-dimensional case with $\mathbf{u}_{\scriptstyle\text{s}}=(\text{u}_{\scriptstyle\text{s}},0)$ as,
\begin{align}
    \lambda_{\scriptscriptstyle\text{intr}} \text{u}_{\scriptstyle\text{s}} &= \nabla_{\!\scriptstyle\text{s}}\big[\gamma_{\scriptstyle 0}(\rho_{\scriptstyle\text{s}}-\rho_{\scriptstyle\text{e}})\big]
    -\nabla_{\!\scriptstyle\text{s}}\cdot\big[\eta(\nabla_{\!\scriptstyle\text{s}} \text{u}_{\scriptstyle\text{s}}+\nabla_{\!\scriptstyle\text{s}} \text{u}_{\scriptstyle\text{s}}^{\scriptscriptstyle\text{T}})\big],~\label{eq:slip_shi}
\end{align}
where the slippage results in surface density $\rho_{\scriptstyle\text{s}}$ 
deviating from the equilibrium $\rho_{\scriptstyle\text{e}}$,
and generates capillary pressure scaled by the surface tension parameter $\gamma_{\scriptstyle 0}$.
The surface density further evolves according to the surface density boundary condition,
\begin{align}
\partial \rho_{\scriptstyle \text{s}}/\partial t+\nabla_{\!\scriptstyle\text{s}}\cdot(\rho_{\scriptstyle\text{s}\,}\text{u}_{\scriptstyle\text{s}})=-\Gamma(\rho_{\scriptstyle\text{s}}-\rho_{\scriptstyle\text{e}}),~\label{eq:density_shi}
\end{align}
where $\Gamma$ is the surface-tension relaxation parameter
and related with the kinetic parameter $\tau$ in~\eqref{eq:conc_qian} of GNBC.
\eqref{eq:density_shi} bears strong analogy to the dynamic wetting boundary condition~\eqref{eq:conc_qian}, 
which is absent from generalized lubrication model.
But, the formulation for surface density is expressed in a non-conserved form
which is not compatible with the thermodynamic framework with the material conservation law~\eqref{eq:density_bedeaux}.}

\textcolor{black}{In the present study, 
we develop a thermodynamically consistent formulation 
of dynamic surface slip for multicomponent fluids, 
incorporating both slip and wetting boundary conditions 
within a sharp-interface framework.
Our framework integrates thermodynamic consistency—following 
the sharp-interface and phase-field formalism~\citep{yue2010sharp,wang2023thermodynamically}—with 
hydrodynamic theory 
grounded in interfacial momentum conservation~\citep{ren2015distinguished,hadjiconstantinou2021atomistic}. 
This combination enables a unified description of slip phenomena 
that captures both thermodynamic and hydrodynamic contributions at the fluid–solid interface.}

\textcolor{black}{A central distinction from
previous slip models lies in 
the treatment of pressure at the fluid–solid interface.
Classical models with the incompressible assumption 
treat pressure as an auxiliary field 
introduced solely to enforce the constraint 
$\nabla \cdot \mathbf{u}=0$, 
and therefore omit it entirely from the slip boundary condition.
In contrast, our adsorption-layer (AL) formulation 
assigns physical volume to the interfacial molecules, 
endowing them with a well-defined pressure 
that must remain thermodynamically coupled to the bulk.
This coupling becomes especially important 
in flows driven by pressure gradients—such as Poiseuille flow—where 
the interfacial pressure must respond to 
the bulk pressure gradient 
rather than being artificially excluded.
Incorporating this pressure coupling 
enables our model to reproduce the confinement-enhanced water slippage 
observed in carbon nanotubes, 
an effect that cannot be explained 
by existing sharp-interface slip models}

\textcolor{black}{The remainder of this paper is organized as follows.
Section 2 introduces the total energy functional 
and derives the dynamic wetting-slip boundary condition 
governing interfacial dynamics through the energy minimization principle. 
The distinctions between our model and previous theories are clarified.
Section 3 presents the validation of our model
and examines flow behaviors influenced 
by the proposed wetting–slip boundary condition,
including classical configurations such as Poiseuille and Couette, and wedge flows.}

\section{Model description}

The governing equations for material and momentum evolution in the bulk fluid 
are not the primary focus of this study; 
they follow the standard Cahn–Hilliard–Navier–Stokes (CHNS) formulation 
and are not reproduced here 
(see Appendix~\ref{sec:GE_bulk} and~\cite{zhang2024multi} for detailed derivation).

\begin{figure}
\includegraphics[width=0.9\textwidth]{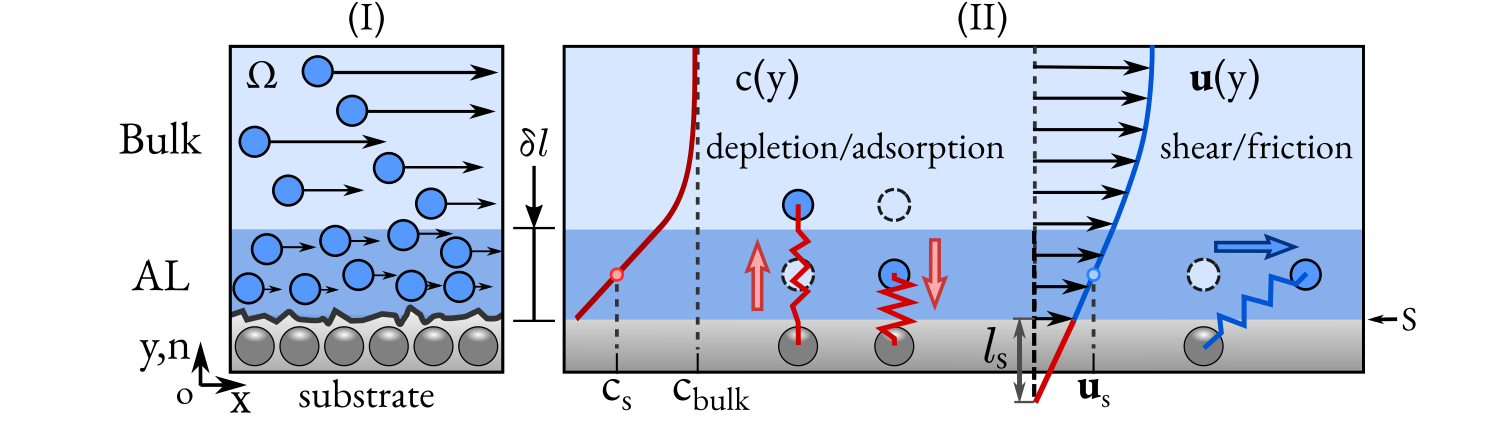}
\centering
\caption{\textcolor{black}{(I) Illustration of fluid passing the solid substrate.
Solid-fluid molecular interactions form the 
adsorption layer (AL) above the solid surface S. 
(II) Solid-fluid molecular interactions perpendicular to S leads to 
surface depletion/adsorption which modifies surface composition $c_{\scriptstyle \text{s}}$ deviating from the bulk fluid.
Interactions parallel to S, such as bond rotation, results in shear/friction
which changes the velocity profile $\mathbf{u}$.}
}
\label{fig:AL}
\end{figure}

In contrast, the present work concentrates on 
the fluid–solid contact region, 
where molecular interactions play a dominant role.
Rather than treating this interface as a mathematically sharp boundary of zero thickness, 
we adopt the concept of adsorption layer (AL), 
analogous to the Helmholtz layer in electrical double-layer (EDL) theory.
In this framework, the fluid–solid interface is modeled as a thin layer of finite physical thickness $\delta l$, 
within which the fluid–solid molecular interactions are significant; 
see figure~\ref{fig:AL}(I).

\subsection{Total energy functional at fluid-solid interface}

\subsubsection{Chemical Free Energy}

Within this interfacial layer, denoted as the adsorption layer (AL), 
the local free energy density comprises contributions from
mixture entropy, mixture enthalpy, and van der Waals interactions~\citep{wang2024wetting,de1985wetting}.
Accordingly, the interfacial free energy density is expressed as,
\begin{align*}
   f_{\scriptstyle \text{w}}(c) =  h_{\scriptstyle \text{w}}(c) -\,T s_{\scriptstyle \text{w}}(c)
    +f_{\scriptstyle \text{vdW}}.
\end{align*}
where $h_{\scriptstyle \text{w}}$
and $s_{\scriptstyle \text{w}}$ 
are the local enthalpy and entropy densities of the fluid mixture in AL.
The absolute temperature is $T$, and
$f_{\scriptstyle \text{vdW}}$ represents the contribution of van der Waals interactions between solid and fluid molecules.
The latter modifies the effective potential energy landscape of the fluid near the solid substrate 
and can be evaluated using a Lennard–Jones–type potential energy density.
Because $f_{\scriptstyle \text{w}}$ depends on the local composition $c(\mathbf{x})$,
the interfacial free energy naturally becomes composition-dependent.
Given that AL is typically much thinner than the bulk region, 
we define an average surface composition, 
similar to the average ion concentration in the EDL of Helmholtz model as
\begin{align*}
    c_{\scriptstyle \text{s}}=\frac{1}{\delta l}\int_{\scriptstyle 0}^{\scriptstyle\delta l}\!\!c(\mathbf{x})_{\scriptscriptstyle \,}\mathrm{d}\mathbf{x}.
\end{align*}

The interfacial free energy density $f_{\scriptstyle \text{w}}$ 
can be locally expanded about the average surface composition $c_{\scriptstyle \text{s}}$,
to obtain a Cahn–Hilliard-type formulation, 
following the methodology developed in our previous work~\citep{cai2024chemo}. 
Consequently, the total chemical free energy functional of the fluid–solid interface is written as
\begin{align*}
    \mathcal{F}_{\scriptstyle \text{s}} =\int_{\scriptstyle \text{S}} \mathrm{d}\text{S}\!\int_{\scriptstyle 0}^{\scriptstyle \delta l}\!\!\!f_{\scriptstyle \text{w}}\big(c(\mathbf{x})\big)\,\mathrm{d}\mathbf{x}
    =\int_{\scriptstyle \text{S}}\Big[f_{\scriptstyle \text{w}}(c_{\scriptstyle \text{s}})+\frac{\kappa_{\scriptstyle \text{s}}}{2}(\nabla c_{\scriptstyle \text{s}})^{\scriptstyle 2}\Big]
    \delta l_{\scriptscriptstyle \,}\mathrm{d}\text{S}
\end{align*}
where the integration is performed over the macroscopic interfacial area S. 
In this expansion, the term proportional to the surface composition gradient 
accounts for the excess energy arising from compositional inhomogeneity along the interface. 
Here, $\kappa_{\scriptstyle \text{s}}$ is the surface gradient energy coefficient, 
quantifying the excess free energy associated with local composition variations along the interface.

As illustrated schematically in figure~\ref{fig:AL}(II),
within the thin layer of thickness $\delta l$ immediately above the uppermost layer of solid molecules, 
fluid molecules experience depletion/adsorption forces 
arising from fluid–solid molecular interactions.
These forces induce a gradual decay (or enrichment) of the local composition 
from the AL towards the bulk fluid region,
as depicted by the red solid line in figure~\ref{fig:AL}(II).

\subsubsection{Kinetic Energy}
In addition to the chemical free energy, 
the fluid within the adsorption layer also carries kinetic energy, 
which is characterized by the height-averaged slip velocity, namely,
\begin{align*}
    \mathbf{u}_{\scriptstyle \text{s}}=\frac{1}{\delta l}\int_{\scriptstyle 0}^{\scriptstyle\delta l}\!\!\mathbf{u}(\mathbf{x})_{\scriptscriptstyle \,}\mathrm{d}\mathbf{x},
\end{align*}
representing the averaged tangential motion of fluid molecules within the AL~\citep{holey2023confinement}.
Consequently, the total kinetic energy contained in AL is expressed as
\textcolor{black}{\begin{align*}
 \mathcal{K}_{\scriptstyle \text{s}} = \frac{1}{2}\int_{\scriptstyle \text{S}}\mathrm{d}\text{S}
\int_{\scriptstyle 0}^{\scriptstyle\delta l}\!\!\!\!\int_{\scriptstyle 0}^{\scriptstyle\delta l} \!\!\mathrm{d}\mathbf{x}\,\mathrm{d}\mathbf{x}^{\scriptstyle \prime}\langle_{\scriptscriptstyle \,}\rho _{\scriptscriptstyle \,}\mathbf u(\mathbf{x}), 
\mathbf{u}(\mathbf{x}^{\scriptstyle \prime})\rangle
=\int_{\scriptstyle \text{S}}\frac{\rho_{\scriptstyle \,}\mathbf{u}_{\scriptstyle\text{s}}^{\scriptstyle\, 2}}{2}\delta l_{\scriptscriptstyle \,}\mathrm{d}\text{S},
\end{align*}
where $\rho$ denotes the local fluid density.
The representative slip velocity $\mathbf{u}_{\scriptstyle \text{s}}$
is defined at the barycentric plane of the adsorption layer
(i.e., at $y=\delta l/2$),
consistent with experimental observations of polymeric wall slip~\citep{cross2018wall}.
This positioning reflects the physical locus 
at which the tangential momentum transfer between the fluid and solid is most effectively mediated.}

\subsubsection{Potential Energy}
For fluid systems, there is also
the potential energy stored within the adsorption layer as
\begin{align*}
 \mathcal{P}_{\scriptstyle\text{s}} =\int_{\scriptstyle \text{S}}\mathrm{d}\text{S}
\int_{\scriptstyle 0}^{\scriptstyle\delta l}\!\!\!\! p(\mathbf{x})\mathrm{d}\mathbf{x}
=\int_{\scriptstyle\text{S}} p_{\scriptstyle\text{s}} \,\delta l\, \mathrm{dS},
\end{align*}
where $p_{\scriptstyle \text{s}}$ denotes 
the average potential energy density in AL.

\subsection{Energy minimization principle}

Following the thermodynamic framework of~\cite{bedeaux1986nonequilibrium,qian2003molecular,bothe2016interface,zhang2024multi},
the governing equations within the adsorption layer
can be systematically derived from the energy minimization principle.
Detailed deductions are documented in the Appendix~\ref{sec:GE_AL}.
The resulting dissipation contains two concurrent minimizing mechanisms
that govern interfacial dynamics.

\subsubsection{Energy Dissipation via Surface Diffusion}
The first mechanism corresponds to surface diffusion, 
which governs interfacial material transport 
and manifests itself through the wetting boundary condition:
\begin{gather}
    \frac{\partial c_{\scriptstyle\text{s}}}{\partial\text{t}}
+ \mathbf{u}_{\scriptstyle \text{s}}\cdot\nabla c_{\scriptstyle\text{s}}
=\nabla\cdot\big(M_{\scriptstyle \text{s}}\nabla\mu_{\scriptstyle\text{s}}\big), \label{eq:GE_AL1}
\end{gather}
where $M_{\scriptstyle\text{s}}$ is the surface diffusion mobility
and $\mu_{\scriptstyle\text{s}}:=\partial f_{\scriptstyle\text{w}}/\partial c_{\scriptstyle\text{s}}-\kappa_{\scriptstyle\text{s}}\nabla^{\scriptstyle 2}c_{\scriptstyle\text{s}}$ 
is the surface chemical potential
defined in the adsorption layer~\citep{dong2021thermodynamics}. 

Here,~\eqref{eq:GE_AL1} takes the Cahn–Hilliard form, 
reflecting the conserved nature of material transport 
within the physically finite AL.
In contrast, earlier sharp-interface formulations~\citep{qian2003molecular,yue2010sharp}, 
treated the interface as a mathematically infinitesimal 
(“ghost”) layer governed by a non-conserved Allen–Cahn type formulation,
such as~\eqref{eq:conc_qian}.
Our model treats the AL as a genuine interfacial region with finite volume, 
consistent with the Helmholtz-layer analogy in the EDL.
Material conservation within this layer becomes essential 
when modeling charge or ion transport, 
motivating our use of the conserved form.
As shown in Appendix~\ref{sec:wbc_revisit}, 
the Allen–Cahn and Cahn–Hilliard forms become mathematically equivalent under specific limits,
ensuring formal consistency with previous models 
while reinstating their physical interpretation.

\textcolor{black}{Our wetting boundary condition~\eqref{eq:GE_AL1} is structurally analogous to 
the surface density conservation equation~\eqref{eq:density_bedeaux},
assuming a linear relation between surface density $\rho_{\scriptstyle\text{s}}$
and composition $c_{\scriptstyle\text{s}}$.
Two distinctions, however, should be emphasized.
First, the convective term in our formulation appears as $\mathbf{u}_{\scriptstyle \text{s}}\cdot\nabla c_{\scriptstyle\text{s}}$,
rather than 
$\nabla\cdot(\rho_{\scriptstyle\text{s}\,}\mathbf{u}_{\scriptstyle \text{s}})$ in their treatment. 
The two expressions are equivalent 
under the incompressibility constraint $\nabla\cdot\mathbf{u}_{\scriptstyle\text{s}}=0$,
ensuring full consistency with continuum mechanics.
Second, we incorporate composition-driven surface diffusion,
which becomes significant in heterogeneous interfacial systems.}

\subsubsection{Energy Dissipation via Convection}
The second dissipation channel arises from momentum convection associated with interfacial slippage.
Averaging the momentum over the adsorption layer yields~\citep{holey2023confinement}:
\begin{gather}
\frac{\mathrm{d}(\rho_{\scriptscriptstyle\,}\mathbf{u}_{\scriptstyle \text{s}})}{\mathrm{d}\text{t}}
 \delta l
=  \nabla \cdot \underline{\underline{\text{T}_{\scriptstyle \text{s}}}}\,\delta l
+\textbf{F}.
 \label{eq:GE_AL2}
\end{gather}
where the stress tensor in AL reads
\begin{equation*}
    \underline{\underline{\text{T}_{\scriptstyle \text{s}}}}= -p_{\scriptstyle \text{s}}\textbf{I} + \kappa_{\scriptstyle\text{s}}\nabla c_{\scriptstyle \text{s}}\otimes\nabla c_{\scriptstyle \text{s}}+\eta (\nabla \mathbf{u}_{\scriptstyle \text{s}}+\nabla \mathbf{u}_{\scriptstyle \text{s}}^{\scriptscriptstyle\text{T}}).
\end{equation*}
The last term on the right-hand side (\textit{rhs}) of  \eqref{eq:GE_AL2} 
corresponds to the Stokes-type drag arising from all external forces.
In this study, $\mathbf{F}$ is identified as the fluid–solid friction,
which plays a central role in controlling interfacial momentum transfer and slip.

\textcolor{black}{To explicitly determine the form of $\mathbf{F}$,
we consider the microscopic origin of friction within AL.
As illustrated in figure~\ref{fig:AL}(II), 
molecular slippage within the adsorption layer (AL) induces local bond rotation and elongation 
between the substrate and adjacent fluid molecules.
Following the hydrodynamic model of~\cite{bocquet1994hydrodynamic},
the rate of energy dissipation due to fluid–solid friction—under a linear response approximation—can be expressed as
\begin{equation*}
    \mathrm{d} E_{\scriptstyle \text{s}}/\mathrm{dt}= -\int_{\scriptstyle 0}^{\scriptstyle\delta l}\!\!\!\int_{\scriptstyle 0}^{\scriptstyle\delta l}\!\!\mathrm{d}\mathbf{x}\,\mathrm{d}\mathbf{x}^{\scriptstyle \prime}\langle\lambda \big(c(\mathbf{x})\big)\mathbf{u}(\mathbf{x}),\mathbf{u}(\mathbf{x}^{\scriptstyle \prime})\rangle = -\lambda(c_{\scriptstyle \text{s}})\,\mathbf{u}_{\scriptstyle \text{s}}^{\scriptstyle 2\,}\delta l,
\end{equation*}
so that the height-averaged friction coefficient 
$\lambda(c_{\scriptstyle \text{s}})$ also depends on the local surface composition $c_{\scriptstyle \text{s}}$.
The mean frictional force acting on the fluid within the AL is thus
\begin{align*}
    \mathbf{F}= (\mathrm{d} E_{\scriptstyle \text{s}}/\mathrm{dt})/\mathbf{u}_{\scriptstyle \text{s}}
= -\lambda(c_{\scriptstyle \text{s}})\,\mathbf{u}_{\scriptstyle \text{s}\,}\delta l.
\end{align*}
Unlike sharp-interface models where $\delta l\rightarrow 0$, 
our finite-thickness formulation explicitly introduces $\delta l$ as a characteristic momentum-dissipation scale,
thereby linking slip attenuation directly to the range of solid–fluid molecular interactions.}

Moreover,~\eqref{eq:GE_AL2} is also expressed as
\begin{gather}
\frac{\mathrm{d}(\rho_{\scriptscriptstyle\,}\mathbf{u}_{\scriptstyle \text{s}})}{\mathrm{d}\text{t}}
=  -\nabla \text{P}_{\scriptstyle \text{s}}
-c_{\scriptstyle \text{s}}\nabla \mu_{\scriptstyle \text{s}}\,
 +\nabla\cdot\big[\eta(\nabla \mathbf{u}_{\scriptstyle \text{s}}+\nabla\mathbf{u}_{\scriptstyle \text{s}}^{\scriptscriptstyle\text{T}})\big]
-\lambda_{\scriptscriptstyle \,}\mathbf{u}_{\scriptstyle \text{s}},
 \label{eq:GE_AL3}
\end{gather}
where the grand pressure is defined as $\text{P}_{\scriptstyle \text{s}}=p_{\scriptstyle \text{s}}+f_{\scriptstyle \text{w}}+\kappa_{\scriptstyle \text{s}}(\nabla c_{\scriptstyle \text{s}})^{\scriptstyle 2}-\mu_{\scriptstyle \text{s}}c_{\scriptstyle \text{s}}$.
The detailed deduction is documented in the Appendix~\ref{sec:GE_AL}.

\subsubsection{Thermodynamic Consistency and Isothermal Approximation}

\textcolor{black}{While the above derivation follows an energy-minimization route,
it remains fully consistent with the entropy principle of~\cite{bedeaux1986nonequilibrium,bothe2022sharp}.
Those studies naturally incorporate interfacial heat 
and entropy fluxes for non-isothermal systems,
whereas our present model focuses on the isothermal limit.
In the following part, we will discuss the connection of both approaches
and their thermodynamically equivalence under specified conditions.}

\textcolor{black}{As deduced in Appendix~\ref{sec:dldt}, 
the total energy dissipation rate at the fluid–solid interface under isothermal condition is
\begin{align*}
\frac{\mathrm{d} \mathcal L_{\scriptstyle\text{s}}(c_{\scriptstyle \text{s}},\textbf{u}_{\scriptstyle \text{s}};T)}{\mathrm{dt}}
=\frac{\mathrm{d}  (\mathcal F_{\scriptstyle\text{s}}+\mathcal K_{\scriptstyle\text{s}}+\mathcal P_{\scriptstyle\text{s}})}{\mathrm{dt}}
= - \!\!\int_{\scriptstyle \text{S}} 
 \big[M_{\scriptstyle \text{s}} (\nabla \mu_{\scriptstyle\text{s}})^{\scriptstyle 2} +  \eta
 \nabla\mathbf u_{\scriptstyle\,\text{s}} :\!\nabla\mathbf u_{\scriptstyle\text{s}}
 +\lambda \,\mathbf u_{\scriptstyle \text{s}}^{\scriptstyle 2}\big]\delta l\,\mathrm{dS}\leq 0.
\end{align*}
Including the entropy $\mathcal{S}$ related to the thermal effects recovers total energy conservation
\begin{align*}
\mathcal L_{\scriptstyle\text{s}}(c_{\scriptstyle \text{s}},\textbf{u}_{\scriptstyle \text{s}},T) + T \mathcal{S}=\text{constant},
\end{align*}
and the entropy production rate becomes
\begin{align*}
\frac{\mathrm{d}\mathcal{S}}{\mathrm{dt}}
=-\frac{1}{T}\frac{\mathrm{d}\mathcal L_{\scriptstyle \text{s}}(c_{\scriptstyle \text{s}},\textbf{u}_{\scriptstyle \text{s}},T) }{\mathrm{dt}}
=-\frac{1}{T}\frac{\mathrm{d}\mathcal L_{\scriptstyle \text{s}}(c_{\scriptstyle \text{s}},\textbf{u}_{\scriptstyle \text{s}};T) }{\mathrm{dt}}
 -\frac{1}{T}\frac{\partial\mathcal L_{\scriptstyle \text{s}} }{\partial T}\frac{\mathrm{d}T}{\mathrm{dt}},
\end{align*}
where $\partial\mathcal L_{\scriptstyle \text{s}}/\partial T$ defines the thermo-chemical potential.
By introducing thermal conductivity $K$, 
we recover Fourier’s heat equation:
\begin{align*}
\mathrm{d}T /\mathrm{dt}&=\nabla\cdot\big[K\nabla(\partial\mathcal L_{\scriptstyle \text{s}}/\partial T)\big]
=\nabla\cdot(\kappa_{\scriptscriptstyle T}\nabla T),
\end{align*}
with the thermal diffusivity $\kappa_{\scriptscriptstyle T}=K\partial^{\scriptscriptstyle 2}\mathcal L_{\scriptstyle \text{s}}/\partial T^{\scriptscriptstyle 2}$.
Hence, the total entropy production rate is
\begin{align*}
\frac{\mathrm{d}\mathcal{S}}{\mathrm{dt}}
=\frac{1}{T}\int_{\scriptstyle \text{S}} 
 \big[\kappa_{\scriptscriptstyle T}(\nabla T)^{\scriptstyle 2} + M_{\scriptstyle \text{s}} (\nabla \mu_{\scriptstyle\text{s}})^{\scriptstyle 2} 
 +  \eta
 \nabla\mathbf u_{\scriptstyle\text{s}} :\!\nabla\mathbf u_{\scriptstyle\text{s}}
 +\lambda \,\mathbf u_{\scriptstyle\text{s}}^{\scriptstyle 2}\big]\delta l\,\mathrm{dS}\geq 0,
\end{align*}
This means when temperature evolution is considered, 
our energy minimization formulation is strictly 
equivalent to the entropy maximization principle of~\cite{bedeaux1986nonequilibrium,bothe2022sharp}.}

\textcolor{black}{For typical liquids, such as water under moderate slip velocities, 
the interfacial heating generated by friction is minimal, 
and the isothermal approximation remains valid.
Using the molecular dynamics estimation~\citep{bui2024revisiting} 
with $\lambda \sim 10^{\scriptstyle 14}\si{kg/(m^{\scriptstyle 3} s)}$ for water confined in carbon nanotube, 
the effective Prandtl number can be estimated as
\begin{equation*}
    \text{Pr} = \frac{\text{Momentum diffusivity by friction force}}{\text{Thermal diffusivity}}
    =\frac{\lambda\delta l^2/\rho}{\kappa_{\scriptscriptstyle T}/(c_{\!\scriptscriptstyle p}\rho)}\approx 7.6 \ (\text{for water}),
\end{equation*}
indicating that momentum diffusion due to friction dominates over thermal diffusion.
Thus, for short-time interfacial processes at ambient conditions, 
the isothermal energy-minimizing framework employed here
is an accurate and physically justified approximation.}

\section{Results and Discussion}

Based on the evolution equations within the adsorption layer, 
\eqref{eq:GE_AL1}
characterizes the dissipation of chemical free energy driven by surface diffusion,
while~\eqref{eq:GE_AL2} governs the dissipation of kinetic energy associated with interfacial slippage.
Both processes are intrinsically coupled through the slip velocity $\mathbf{u}_{\scriptstyle \text{s}}$.

Regarding the momentum equation in the AL, 
 \eqref{eq:GE_AL2} now serves as a dynamic momentum evolution equation 
describing the development and dissipation of the slip velocity 
under the combined influence of viscous, capillary, and frictional stresses.
From a computational perspective, 
this modification can be readily incorporated into existing Navier–Stokes solvers by embedding AL 
as a thin interfacial region,
where the friction term is explicitly introduced without needs for major structural alterations to the solver.

Our AL-based momentum equation can also recover
the generalized Navier boundary condition (GNBC),
when the following four assumptions are satisfied: 
(i) at steady-state condition $\partial \mathbf{u}_{\scriptstyle \text{s}}/\partial \text{t}=0$ 
and (ii) negligible fluid inertia $\rho\mathbf{u}_{\scriptstyle \text{s}}\cdot\nabla\mathbf{u}_{\scriptstyle \text{s}}=0$,
 \eqref{eq:GE_AL2} reduces to the local force balance:
\begin{align}
\lambda\,\mathbf{u}_{\scriptstyle\text{s}\,}\delta l
=\eta_{\scriptscriptstyle\,}
(\nabla \mathbf{u}+\nabla\mathbf{u}^{\scriptscriptstyle\text{T}})\cdot \mathbf{n}
+ [-\nabla \text{P}_{\scriptstyle \text{s}} -c_{\scriptstyle\text{s}}
\nabla\mu_{\scriptstyle\text{s}}
+\nabla_{\!\scriptstyle \text{s}}\cdot(\eta
\nabla_{\!\scriptstyle \text{s}}\mathbf{u}_{\scriptstyle\text{s}})]\delta l.\label{eq:v_stable}
\end{align}
Further simplifications are achieved 
under (iii) vanishing grand pressure gradient $\nabla \text{P}_{\scriptstyle \text{s}} = 0$, 
and (iv) negligible viscous stress for fluid molecules in AL,
namely, $\nabla_{\!\scriptstyle \text{s}}\cdot
(\eta\nabla_{\!\scriptstyle \text{s}}\mathbf{u}_{\scriptstyle \text{s}})$,  
\eqref{eq:v_stable} 
recovers the generalized Navier boundary condition 
 \eqref{eq:slip_qian}.

It is evident, however, 
that these criteria are not universally satisfied.
For instance, criterion (iii) fails for single-phase Poiseuille flow,
while in multiphase systems—where interfacial curvature 
and viscosity contrasts exist—criterion (iv) does not hold 
for typical Couette or Stokes flow configurations.
Therefore, our model extends the validity of the GNBC 
by explicitly retaining the viscous and capillary contributions in  \eqref{eq:v_stable},
providing a physically consistent and energetically grounded description of the interfacial slip.

In the subsequent analysis, 
we apply the present slip model to several representative flow configurations.
To highlight the dissipation mechanisms at fluid-solid interfaces with slippage,
we employ characteristic material properties of water, 
namely its viscosity and interfacial tension.
Under these conditions, the system exhibits a low Reynolds number ($\text{Re}\ll 1$),
ensuring negligible inertial effects.
This permits a detailed examination of steady-state solutions
that describe interfacial slip and frictional dissipation,
enabling a direct comparison with classical theories.
Because the focus of this work is on the theoretical formulation 
and analytical understanding of slip phenomena,
we do not include numerical simulations here.

\subsection{Model Validation}
\textcolor{black}{In this part, we consider a simple single-phase flow with confinement
to validate the proposed model at steady states 
with the negligible fluid inertia,
so that $\mathrm{d}\text{u}_{\scriptstyle \text{s}}/\mathrm{dt}=0$.
Under this condition, the slip boundary condition~\eqref{eq:v_stable} is simplified as
\begin{align*}
    \lambda\, \mathbf{u}_{\scriptstyle \text{s}\,}\delta l
=\eta_{\scriptscriptstyle \,}\nabla\mathbf{u}\cdot\mathbf{n}
-\delta l_{\scriptscriptstyle \,}\nabla \text{P}_{\scriptstyle \text{s}}.
\end{align*}}
\textcolor{black}{A key distinction from the classical Navier slip boundary condition (NBC) 
is the presence of the grand pressure-gradient term $\nabla \text{P}_{\scriptstyle \text{s}}$,
scaled by the adsorption layer thickness $\delta l$, 
which explicitly enters the interfacial force balance.
Such a term has been largely neglected in previous sharp-interface 
treatments~\citep{snoeijer2006free,chan2020cox,bui2024revisiting}. 
By explicitly resolving the finite-thickness adsorption layer,
our formulation restores a complete local force balance 
acting on the interfacial region,
including both the tangential (parallel) 
and normal components relative to the solid substrate.}

\textcolor{black}{For simple incompressible flows, 
the force balance in normal direction implies 
\begin{align*}
   ( -\nabla \text{P}_{\scriptstyle \text{s}}-c_{\scriptstyle \text{s}}\nabla\mu_{\scriptstyle \text{s}})\cdot \mathbf{n}=0.
\end{align*}
Due to the thermodynamic equilibrium for component,
$\nabla\mu_{\scriptstyle \text{s}}\cdot \mathbf{n}=0$,
so that
\begin{align}
    \nabla \text{P}_{\scriptstyle \text{s}}\cdot \mathbf{n}=0,~\label{eq:p_couple}
\end{align}
per se $\text{P}_{\scriptstyle \text{s}}=\text{P}$,
indicating that the grand pressure transmitted from the bulk 
must also act upon the fluid molecules within the AL.
Consequently, a pressure gradient persists inside the AL, 
contributing to fluid motion 
and playing a crucial role in maintaining mechanical equilibrium.
This interfacial condition naturally arises from the energy minimization principle
and remains consistent with the entropy production 
framework~\citep{bedeaux1986nonequilibrium,bothe2022sharp}.}

\subsubsection{One-dimensional Poiseuille flow}

\begin{figure}
\includegraphics[width=0.9\textwidth]{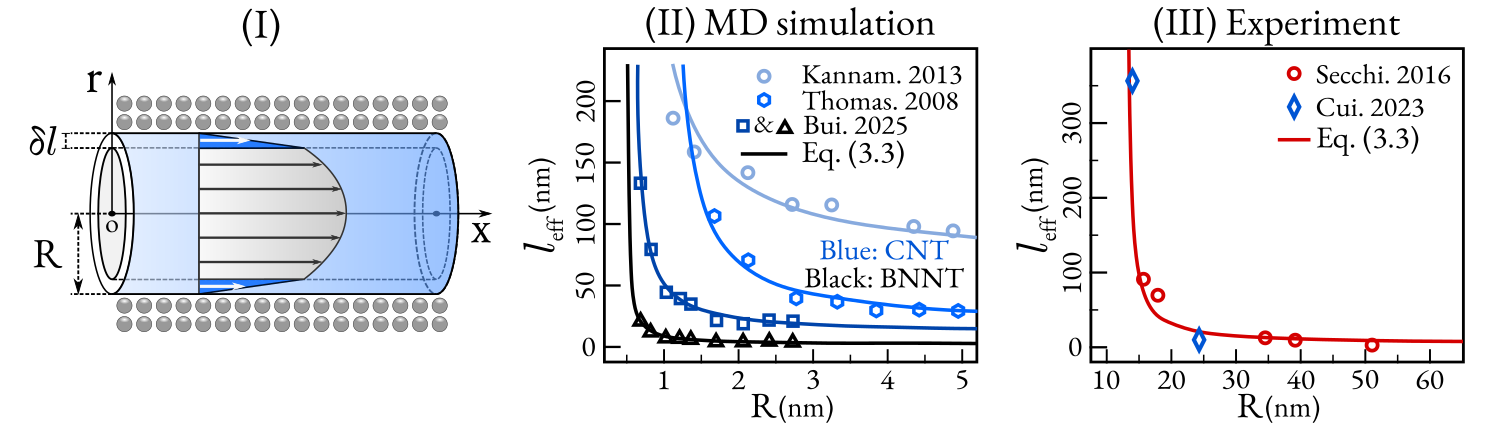}
\centering
\caption{\textcolor{black}{
(I) One-dimensional Poiseuille flow through a tube with radius R.
(II) Calculated effective slip length $l_{\scriptstyle \text{eff}}$ of water 
in nanotube with  \eqref{eq:slip_l} shows R-dependent decays,
compared with previous molecular dynamic simulations.
CNT: carbon nanotube, BNNT: boron nitride nanotube,
for both $\delta l = 0.5\si{nm}$.
(III) Enhanced experimental slip length of KCl (potassium chloride) aqueous 
solution ($10^{\scriptstyle -3}$M) in CNT deviates from MD,
indicating the larger $\delta l \approx 13\si{nm}$
associated with the Debye length of KCl solution.}
}
\label{fig:poiseuille}
\end{figure}
To further elucidate the physical implications,
we first examine the simplest case:
a steady Poiseuille flow under confinement in a cylindrical pipe of radius R,
as illustrated in figure~\ref{fig:poiseuille}(I)(i).
The velocity field $\mathbf{u}=(\text{u},0)$ is axisymmetric and unidirectional along the axial direction x.
In cylindrical coordinates,
the flow satisfies the Stokes equation in the bulk fluid 
and the force balance in the adsorption layer (AL) described 
by~\eqref{eq:GE_AL2} in AL, namely:
\begin{gather*}
   \left\{
    \begin{array}{ll}\partial_{\scriptstyle \text{x}} \text{P}=\nabla\cdot(\eta\nabla\text{u}),&\text{in r} \in [0, \text{R}-\delta l],\\[0.3em]
   \lambda\, \text{u}_{\scriptstyle \text{s}\,}\delta l=-\eta_{\scriptscriptstyle \,}\partial_{\scriptstyle \text{r}}\mathrm{u}-\delta l_{\scriptscriptstyle \,}\partial_{\scriptstyle \text{x}}\text{P}_{\scriptstyle \text{s}}, \quad  &\text{in r}=\text{R}-\delta l/2.
   \end{array} \right.
\end{gather*}

The first equation represents the Stokes flow in the bulk,
while the second expresses the local momentum balance within the adsorption layer (AL).
Notably, conventional sharp-interface models~\citep{chan2020cox,bui2024revisiting}
neglect the grand pressure-gradient term in the slip boundary condition,
thereby decoupling interfacial pressure from that of the bulk.
In contrast, following the variational formulation 
of~\citet{bedeaux1986nonequilibrium,qian2006variational},
our framework explicitly couples the adsorption-layer pressure 
with that of the bulk fluid, 
ensuring thermodynamic and hydrodynamic consistency for AL and the bulk fluid.
Due to the pressure coupling between AL and 
the bulk fluid~\eqref{eq:p_couple}, 
we have $\partial_{\scriptstyle\text{x}} \text{P}_{\scriptstyle\text{s}}=\partial_{\scriptstyle\text{x}}\text{P}$.

\textcolor{black}{Accounting for this newly added pressure-coupling term, 
the steady-state velocity profile now becomes
\begin{align*}
    \text{u}(\text{r})=\frac{\mathrm{d}_{\scriptstyle \text{x}}\text{P}}{4\eta}\big[\text{r}^2-(\text{R}-\delta l)^2
    -4(\text{R}+\delta l)l_{\scriptstyle \text{s}}\big],
\end{align*}
where $l_{\scriptstyle \text{s}}=\eta/(\lambda\delta l)$ 
is the intrinsic Navier slip length, 
characterizing the frictional coupling 
between the wall and the adjacent fluid molecules.
This leads naturally to a confinement-dependent effective slip length,
defined as
\begin{align}
    l_{\scriptstyle \text{eff}} = 
    \frac{\,\text{u}_{\scriptstyle \text{s}}}{\ \partial_\text{r} \text{u}\ }
    =\Big(1+\frac{2\delta l}{\text{R}-\delta l}\Big)\,l_{\scriptstyle\text{s}},~\label{eq:slip_l}
\end{align}
which explicitly depends on both the pipe radius and the AL thickness.}

\textcolor{black}{Figure~\ref{fig:poiseuille}(II) 
compares~\eqref{eq:slip_l} with molecular dynamics (MD) simulations 
of water flow in carbon nanotubes (CNTs) 
and boron nitride nanotubes (BNNTs)~\citep{thomas2008reassessing,kannam2013fast,bui2024revisiting}. 
The characteristic solid–fluid interaction range is taken as $\delta l=0.5\si{nm}$.
For sufficiently large pipes $(\text{R}\gg \delta l)$, 
 \eqref{eq:slip_l} asymptotically recovers the classical Navier slip length,
$l_{\scriptstyle \text{eff}}\approx l_{\scriptstyle\text{s}}$,
where confinement effects are negligible.
At smaller radii, however, our model predicts a pronounced slip enhancement with decreasing R,
in well agreement with MD data.
Importantly, our framework also captures the significant difference 
in slip length between CNTs and BNNTs—despite their nearly identical macroscopic 
wettability—by attributing it to variations in the interfacial friction coefficient 
$\lambda$.
Such variations arise from molecular-scale features related to the fluid-solid intermolecular bond, such as polarization and surface corrugation, 
rather than from contact angle differences alone.}

\textcolor{black}{We further apply~\eqref{eq:slip_l} 
to interpret the extraordinarily large slip lengths
observed for electrolytic flows in CNTs,
which exceed the values for pure water by several orders of magnitude~\citep{secchi2016massive,cui2023enhanced}.
In these experiments, a $10^{\scriptstyle -3}$M KCl (potassium chloride) solution is used.
Without altering other parameters, 
replacing the characteristic interfacial length 
by the Debye screening length $\delta l = 13\,\si{nm}$—representing the ion–solid interaction range—yields theoretical predictions that agree quantitatively 
with experimental results; 
see figure~\ref{fig:poiseuille}(III).
This demonstrates that electrostatic interactions 
within the Helmholtz (or Debye) layer 
can effectively enlarge the hydrodynamic boundary-layer thickness, 
thus amplifying the apparent slip length.
Accordingly, the present model bridges the molecular and continuum descriptions of interfacial slippage, 
providing a unified interpretation for both neutral and electrolytic confined flows.}

\textcolor{black}{Finally, we note that such confinement-induced slip enhancement cannot be reproduced by the classical Navier slip condition,
which neglects the thermodynamic coupling between interfacial and bulk pressures.
Although~\cite{qian2006variational} introduced a pressure term 
in their variational force balance, 
their formulation only addressed the tangential component along the substrate,
as the sharp-interface assumption precludes defining the interfacial pressure 
or its normal balance.
In contrast, our adsorption-layer framework restores a physically meaningful interfacial region 
populated by fluid molecules that remain thermodynamically coupled to the bulk.
As a result, while the friction coefficient $\lambda$ 
remains an intrinsic property of the solid–fluid bonds,
the apparent slip length becomes confinement-dependent 
due to the pressure coupling.
This interpretation differs fundamentally from the confinement-dependent $\lambda$
hypothesis proposed by~\cite{bui2024revisiting}.
We emphasize that $\lambda$ reflects the intrinsic strength of intermolecular bonding rather than flow geometry;
its apparent variation arises indirectly through the pressure-mediated interaction between the bulk fluid and the AL.
}

\subsubsection{Two-dimensional Poiseuille flow}
When the curvature of the geometry is neglected—for example, 
in two-dimensional Poiseuille flow between parallel plates separated by a distance $2$H—the effective slip length takes
\begin{align}
    l_{\scriptstyle \text{eff}} = \Big(1+\frac{\delta l}{\text{H}-\delta l}\Big)\,l_{\scriptstyle\text{s}}.~\label{eq:slip_2d}
\end{align}
This expression indicates that as the confinement becomes stronger $\text{H}\rightarrow \delta l$, 
the slip length increases due to the increased contribution of interfacial slippage relative to the total height of the channel.
This confinement-induced enhancement was not captured in MD simulations~\citep{bui2024revisiting},
where a geometry-independent slip length is prescribed as a constant value.

Based on  \eqref{eq:slip_2d},
the corresponding effective interfacial friction coefficient follows directly from the Navier definition by  \eqref{eq:L_s},
\begin{align}
    \lambda_{\scriptstyle \text{eff}} 
    =\frac{\eta}{l_{\scriptstyle \text{eff}}}
    = \frac{\text{H}-\delta l}{\text{H}}
    \lambda_{\scriptstyle \text{intr}},~\label{eq:lambda_eff}
\end{align}
where $\lambda_{\scriptstyle \text{intr}}=\eta/l_{\scriptstyle \text{s}}$ 
denotes the intrinsic interfacial friction coefficient.
In the limit $\text{H}\rightarrow\infty$, 
 \eqref{eq:lambda_eff} converges to the bulk value, as expected.
This scaling differs from the prediction of the classical hydrodynamic theory (CHT)~\citep{bui2024revisiting}, which yields
\begin{align}
    \lambda_{\scriptstyle \text{eff,CHT}} = \frac{6_{\scriptscriptstyle \,}\eta}{\text{H}+3l_{\scriptstyle\text{s}}},~\label{eq:lambda_eff_cht}
\end{align}
implying that the effective friction vanishes for very large channel heights ($\text{H}\rightarrow\infty$).
Our formulation, in contrast, captures the finite friction resistance sustained by the interfacial layer even in wide channels, 
highlighting the physical role of the adsorption layer thickness $\delta l$ in modulating momentum transfer.

\subsubsection{Couette Flow}
We next consider the two-dimensional Couette flow 
confined between two parallel plates separated by a distance H.
The upper plate moves with constant velocity $\text{U}_0$ at $\text{h}=\text{H}$, 
while the lower plate remains stationary at $\text{h}=0$.
Within the bulk, the velocity field satisfies the Stokes equation 
with a uniform shear rate, 
whereas the momentum balance in the adsorption layer (AL) introduces 
a velocity jump governed by the frictional coupling at each wall.
The analytical solution of the steady velocity profile reads
\begin{align}
    \text{u}(\text{h}) = \text{U}_0\,\frac{\text{h}+2l_{\scriptstyle \text{s}}-\delta l}{\text{H}+4l_{\scriptstyle \text{s}}-2\delta l},\label{eq:u_couette}
\end{align}
which reduces to the classical linear Couette profile in the limit
$\delta l\rightarrow 0$, 
confirming consistency with the no-slip condition.
The corresponding effective slip length evaluated at the wall interface is
$l_{\scriptstyle \text{eff}} = l_{\scriptstyle \text{s}}$,
indicating that, unlike the Poiseuille configuration, 
the Couette flow does not exhibit geometric dependence of the slip length on the confinement spacing H.

\subsection{Surface Depletion/Adsorption Effect}

\textcolor{black}{In multicomponent fluid systems, 
the solid substrate may preferentially adsorb or deplete one component, 
leading to a surface composition that deviates from the bulk mixture.
This compositional gradient along the substrate normal direction induces a spatially varying viscosity field,
thereby altering the local hydrodynamics within AL.
Considering the normal force balance in AL, we have
\begin{align*}
    \nabla \text{P}_{\scriptstyle \text{s}}\cdot\mathbf{n}=c_{\scriptstyle \text{s}}\nabla \mu_{\scriptstyle \text{s}}\cdot\mathbf{n}=0.
\end{align*}
whose solution gives rise to the steady-state concentration profile 
following an exponential relaxation,
depending on the surface depletion/adsorption~\citep{wang2023thermodynamically}.
Thus, the viscosity distribution in the 1D Poiseuille flow can be approximated as
\begin{align*}
    \eta(\text{r}) 
=1+\eta_{\scriptstyle 0}e^{\scriptstyle \text{R}/\delta l}\cosh(\text{r}/\delta l),
\end{align*}
where $\eta_{\scriptscriptstyle 0}$ characterizes the magnitude of viscosity modulation induced by surface adsorption or depletion.
Substituting this expression into the momentum balance yields the steady velocity profile,
\begin{align*}
    \text{u}(\text{r}) = -\frac{1}{2}\int_{\scriptstyle 0}^{\scriptstyle\text{r}}\!\!\frac{s\,\mathrm{d}\text{P}/\mathrm{d}x+C_1}{1+\eta_{\scriptstyle 0}e^{\scriptstyle \text{R}/\delta l}\cosh(s/\delta l)}\,\mathrm{d}s+C_2,
\end{align*}
where the integration constants $C_1, C_2$ are decided by slip boundary conditions at wall.}

\begin{figure}
\includegraphics[width=0.9\textwidth]{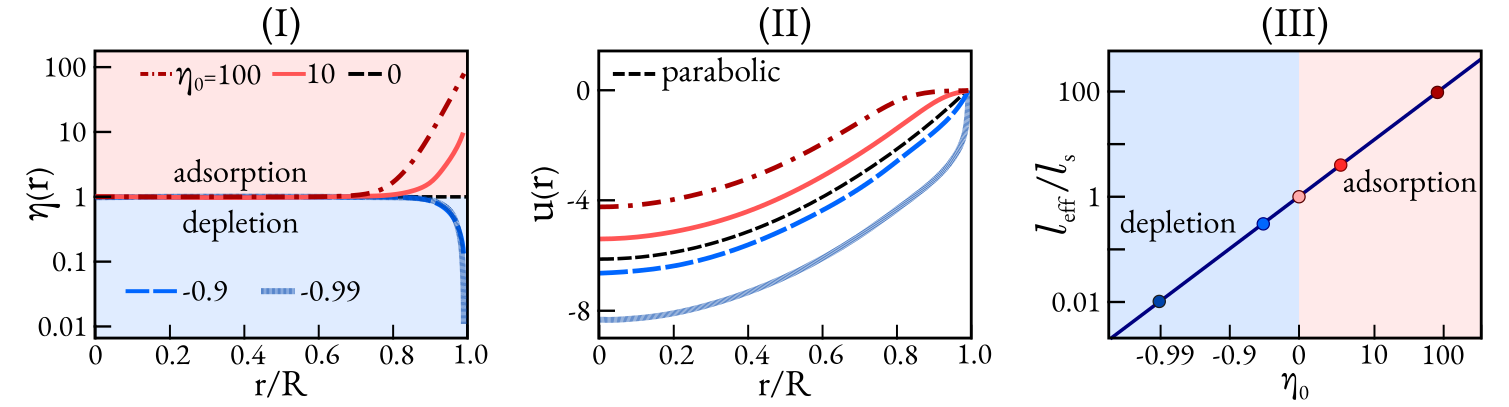}
\centering
\caption{\textcolor{black}{One-dimensional Poiseuille flow with surface depletion/adsorption.
Tube radius $\text{R}=50$, 
characteristic length $\delta l=0.5$,
friction coefficient $\lambda=20$.
(I) Depletion/adsorption induced viscosity distribution 
$\eta(\text{r})=1+\eta_{\scriptstyle 0}e^{\scriptstyle \text{R}/\delta l}\cosh(\text{r}/\delta l)$.
$\eta_{\scriptstyle 0}>0$: adsorption increases viscosity;
$\eta_{\scriptstyle 0}<0$: depletion reduces viscosity.
(II) Flow profile modified by surface depletion/adsorption.
Black line: parabolic profile without depletion/adsorption.
(III) Effective slip length $l_{\scriptstyle \text{eff}}$ changing with depletion/adsorption; dots:  \eqref{eq:slip_l},
line: linear relation.}
}
\label{fig:poiseuille2}
\end{figure}

\textcolor{black}{By varying $\eta_{\scriptstyle 0}$, 
the resulting viscosity and velocity distributions are depicted in figure~\ref{fig:poiseuille2}(I)-(II).
For $\eta_{\scriptstyle 0}>0$, the local viscosity near the wall increases due to surface adsorption, 
producing a pronounced suppression of the near-wall velocity.
Conversely, $\eta_{\scriptstyle 0}<0$, 
corresponds to surface depletion, 
reducing viscosity near the wall 
and thereby enhancing the flow velocity within the AL region.}

\textcolor{black}{Interestingly, when evaluating the effective slip length from the velocity gradient using Navier’s definition of  \eqref{eq:slip_l},
the computed $l_{\scriptstyle \text{eff}}$ appears larger for the more viscous (adsorptive) case, 
as shown in figure~\ref{fig:poiseuille2}(III).
This seemingly counterintuitive trend arises because a steeper velocity gradient 
$\partial\text{u}/\partial \text{r}$ 
develops near the fluid-solid interface 
when the local viscosity increases, 
thereby inflating the apparent slip length defined through this ratio.}

\textcolor{black}{This observation reveals a fundamental limitation of using “slip length” as a universal descriptor of interfacial hydrodynamics.
The parameter $l_{\scriptstyle \text{eff}}$ does not always reflect a genuine slip phenomenon 
but rather encodes the combined effects of interfacial friction, local viscosity variation, and flow geometry.
A more physically meaningful characterization should therefore rely on the explicit interfacial parameters—namely, the local viscosity $\eta(c, \text{r})$
and the friction coefficient $\lambda$—instead of prescribing or inferring a single slip length.
Such an approach avoids the ambiguity of interpreting
$l_{\scriptstyle \text{eff}}$ in complex or compositionally heterogeneous systems.}

\begin{figure}
\includegraphics[width=0.9\textwidth]{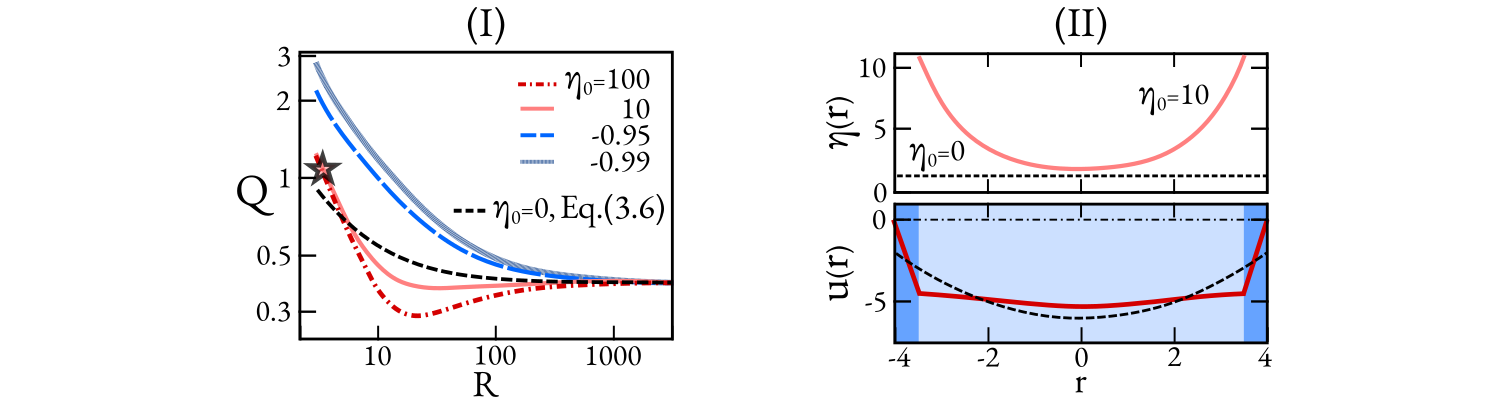}
\centering
\caption{\textcolor{black}{(I) Flow rate of one-dimensional Poiseuille flow with slip influenced by surface depletion/adsorption
for different tube radius R.
The black dashed line: standard $Q_{\scriptstyle \text{s}}$ by  \eqref{eq:qs}.
(II) Viscosity $\eta$ and flow velocity $\text{u}(\text{r})$ 
deviate from the standard Poiseuille flow without surface adsorption (dashed black lines),
for small confinement with $\text{R}=4$, corresponding to the star symbol in (I).
The dark blue and light blue regions denote AL and bulk fluid, respectively.}
}
\label{fig:poiseuille3}
\end{figure}

\textcolor{black}{As the velocity profile $\mathrm{u}(\text{r})$ is modified by the local viscosity variations caused by surface adsorption or depletion, 
the volumetric flow rate through the cylindrical conduit is obtained by integrating the velocity field,
\begin{equation*}
    Q=\int_{\scriptstyle 0}^{\scriptstyle \text{R}}\!\!2\pi \text{r}_{\scriptscriptstyle \,} \text{u}(\text{r})_{\scriptscriptstyle \,}\mathrm{d}\text{r}.
\end{equation*}
This expression deviates from the standard Poiseuille flow rate, $Q_{\scriptstyle \text{s}}$, 
associated with a uniform viscosity and Navier slip boundary condition~\citep{lauga2003effective},
\begin{equation}
    Q_{\scriptstyle \text{s}}=Q_{\scriptstyle 0}\Big(1+\frac{4_{\scriptscriptstyle \,}l_{\scriptstyle \text{s}}}{\text{R}}\Big),\label{eq:qs}
\end{equation}
where $Q_{\scriptstyle 0}=-\pi\text{R}^{\scriptstyle 4}\mathrm{d}_{\scriptstyle \text{x}} \text{P}/(8\eta)$ 
is the classical no-slip flow rate.
 \eqref{eq:qs} corresponds to a special case of our model with $\eta_{\scriptstyle 0}=0$, 
i.e., uniform composition without adsorption layer effects.}

\textcolor{black}{Figure~\ref{fig:poiseuille3}(I) compares the normalized flow rate $Q/Q_{\scriptstyle \text{s}}$
under various conditions of surface depletion/adsorption.
When surface depletion leads to a local viscosity reduction near AL ($\eta_{\scriptstyle 0}<0$), 
the flow rate increases monotonically relative to $Q_{\scriptstyle \text{s}}$ 
and gradually approaches the standard Poiseuille value as the tube radius increases.
Conversely, for $\eta_{\scriptstyle 0}>0$, 
where adsorption enhances the viscosity near the wall, 
the flow rate exhibits a non-monotonic dependence on the confinement radius R.
At small confinements $(\text{R}=4)$, 
as shown in figure~\ref{fig:poiseuille3}(II), 
the velocity profile becomes flattened near the channel centerline 
and elevated within the adsorption layer, 
deviating markedly from the classical parabolic profile (black dashed line).
This counterintuitive enhancement of flow at small 
R arises because, within the adsorption layer, the viscous drag force dominates over the interfacial frictional resistance, thereby producing an increased slip velocity despite a locally higher viscosity.
Although in all configurations the prescribed effective slip length $l_{\scriptstyle \text{s}}=1.0$ is held constant, 
the modified near-wall viscosity distribution alters the apparent flow enhancement, 
demonstrating that the same $l_{\scriptstyle \text{s}}$ 
can correspond to different global transport rates.}

\textcolor{black}{Consequently, using the standard relation in  \eqref{eq:qs} to infer slip length from measured flow rates 
may lead to significant over- or under-estimation, 
depending on the degree of surface adsorption or depletion.
This sensitivity highlights the non-universality of the slip length concept 
when viscosity varies spatially near the interface.
MD simulations of pure water slippage indeed 
reveal a viscosity deviation from the bulk value 
within several molecular diameters of the solid surface.
In experimental studies involving multicomponent fluids, 
even when perfectly smooth substrates are used, 
such near-wall viscosity variations can substantially affect the hydrodynamic response.
Hence, the conventional practice of determining $l_{\scriptstyle \text{s}}$ via  \eqref{eq:qs} 
should be re-examined; 
accurate characterization of slip requires accounting for composition-dependent viscosity fields 
and adsorption layer dynamics.}

\subsection{Wedge flow}
\begin{figure}
\includegraphics[width=0.9\textwidth]{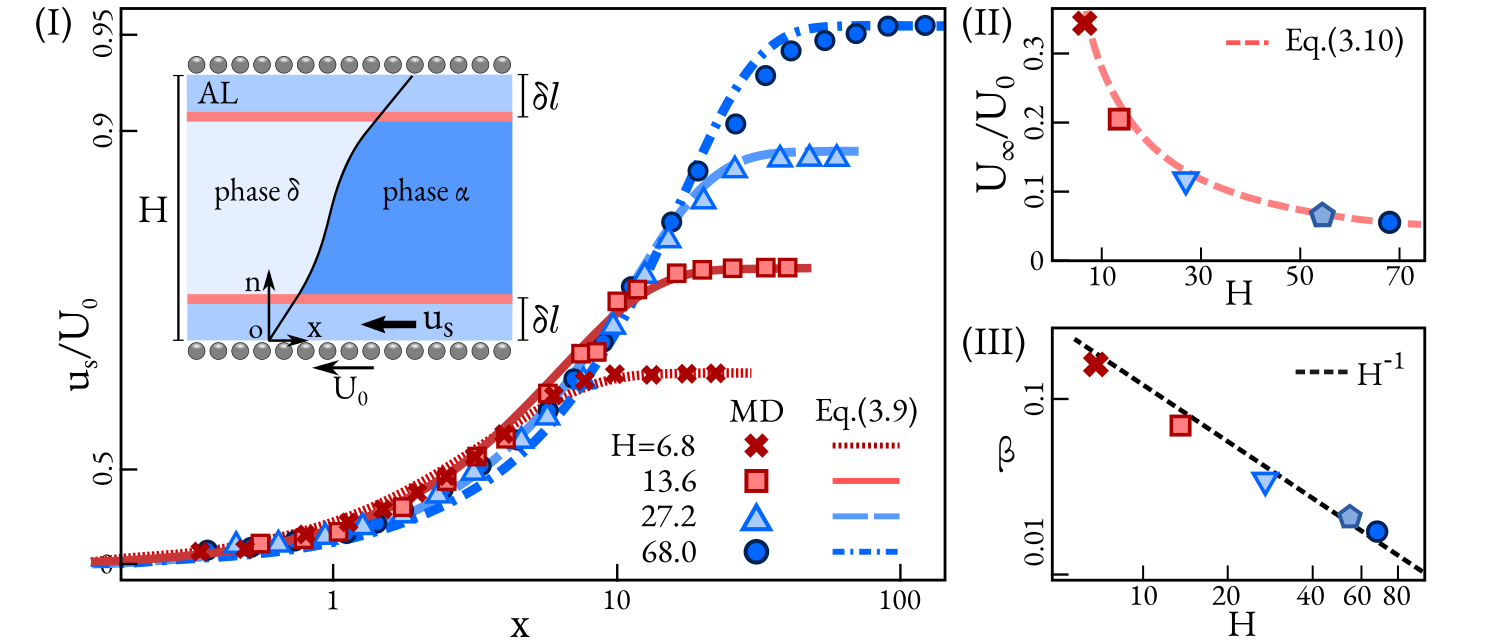}
\centering
\caption{
(I) Flow velocity $\text{u}_{\scriptstyle \text{s}}$ in AL from the triple junction $x=0$.
Open-colored symbols: $\text{u}$ by molecular dynamics simulation 
from~\citet{qian2006variational}.
Colored curves: theory with  \eqref{eq:triple2}.
Inset: flow setup.
(II) Far field flow velocity $\text{U}_{\scriptstyle \infty}$ in AL
fulfills Couette flow with  \eqref{eq:triple3}.
(III) Fitting parameter $\beta$ scaling with the plane spacing H.
Colored symbols in (II)(III) correspond to the markers in (I).}
\label{fig:Fit}
\end{figure}

\textcolor{black}{In the previous section, 
we analyzed single-phase confined flows, 
highlighting how slippage arises from the coupling 
between the bulk fluid and the adsorption layer.
There, the pressure-gradient term and surface composition 
were shown to play crucial roles through their thermodynamic equilibrium
in the direction normal to the fluid-solid interface, 
indirectly modulating the slip velocity.
We now extend our analysis to 
a two-phase configuration—the $\alpha-\delta$ system—forming 
a wedge-shaped flow near the moving contact line, 
as illustrated in the inset of figure~\ref{fig:Fit}(I).}

\textcolor{black}{First analyzed by~\citet{huh1971hydrodynamic},
they assumes that the wedge flow is constraint by the no-slip boundary condition,
while neglecting local force balance near the contact line. 
Solving the Stokes equation in polar coordinates $(r,\phi)$,
their model leads to the well-known Huh–Scriven paradox, 
in which the viscous stress diverges as $\text{r}\rightarrow 0$.
Later studies resolved this paradox 
by introducing microscopic interfacial effects.
Similarly,~\cite{hocking1977moving,chan2020cox}
employed the Navier slip condition 
to regularize the stress singularity,
while~\citet{shikhmurzaev1993moving} introduced 
a surface composition–dependent slip model 
to achieve a comparable effect.
\citet{qian2006variational} and~\citet{yue2010sharp} 
showed numerically that the paradox disappears 
once the liquid–liquid interface is treated as a diffuse region 
with finite thickness, 
ensuring that $\text{r}$ never strictly vanishes.
Building on these developments, 
we revisit the velocity distribution near the triple junction 
by incorporating the viscous stress term
$\nabla_{\!\scriptstyle\text{s}}\cdot(\eta\nabla_{\!\scriptstyle\text{s}}\mathbf{u}_{\scriptstyle \text{s}})\delta l$ in~\eqref{eq:v_stable}
that naturally arises from our adsorption-layer formulation—yet 
has been overlooked in prior treatments.
Using the parameters of~\citet{qian2006variational}, 
we compare our analytical predictions against 
molecular dynamics results for validation.}

\subsubsection{Velocity Profile in Adsorption Layer}
Following the setup of~\citet{qian2004power,qian2006variational}, 
we consider a two-dimensional Couette-type flow 
where the lower substrate moves 
with velocity $\text{U}_{\scriptstyle 0}$.
Both fluid phases are assumed to have identical viscosities, 
such that the velocity field is symmetric about $\text{x}=0$. 
Hence, only the right-hand phase ($\alpha$, with $\text{x}>0$) is analyzed.
In this configuration, no externally imposed pressure gradient exists 
as in Poiseuille flow.
Instead, the Laplace pressure gradient $-\nabla p$ 
is balanced by the thermodynamic driving force
$-c_{\scriptstyle \text{s}}\partial_{\scriptstyle \text{x}}\mu_{\scriptstyle \text{s}}$ 
at steady state, particularly when the slip velocity remains small.
Thus, the grand pressure gradient $\nabla \text{P}_{\scriptstyle \text{s}}$ becomes negligible.
Under these assumptions, the momentum balance in the AL \eqref{eq:v_stable} reduces to
\begin{align}
\lambda(\text{u}_{\scriptstyle \text{s}}-\text{U}_{\scriptscriptstyle 0})\delta l = \eta
\frac{\partial^{\scriptstyle 2} \text{u}_{\scriptstyle \text{s}}}{\partial \text{x}^{\scriptstyle 2}}\delta l
+\eta\frac{\partial \text{u}}{\partial \text{n}}.
\label{eq:triple0}
\end{align}

Far from the contact line ($\text{x}\gg 1$), 
the viscous stress term
$\eta\partial_{\scriptstyle xx}\text{u}$ becomes negligible,
and the flow transitions into a quasi–Couette regime.
Near the contact line, however, the slip velocity decreases sharply 
and the fluid within the AL becomes nearly stagnant,
indicating a finite viscous stress contribution that cannot be ignored.
This slip velocity gradient in substrate tangent direction is also scaled
by AL thickness $\delta l$,
which reveals that friction and viscous coupling act over a finite length scale, rather than at a singular boundary.

At the far field $(x\rightarrow\infty)$,
the slip velocity approaches the asymptotic value
$\text{u}_{\scriptstyle \text{s}}(\infty)=\text{U}_{\scriptstyle\infty}$,
which is determined by the Couette flow expression~\eqref{eq:u_couette}.
Assuming that the near-contact-line flow is generated by this Couette motion,
we approximate the viscous traction term in~\eqref{eq:triple0} as
$\eta\partial_{\scriptstyle n}\text{u} \simeq \nu(\text{u}_{\scriptstyle \text{s}}-\text{U}_{\scriptstyle\infty})$,
where $\nu$ is a geometry-dependent scaling parameter extracted from MD simulations.
Substituting into~\eqref{eq:triple0} gives an exponentially decaying slip-velocity profile within the AL:
\begin{align}
\text{u}_{\scriptstyle \text{s}}(\text{x})
&=(\text{U}_{\scriptstyle \text{c}}-\text{U}_{\scriptstyle\infty})\,\text{e}^{-\beta \text{x}}+\text{U}_{\scriptstyle \infty},
\label{eq:triple2}
\end{align}
where $\text{U}_{\scriptstyle \text{c}}$ denotes the slip velocity at the contact line.
The decay rate $\beta=\sqrt{(\lambda+\nu)/\eta}$ depends on the flow geometry 
and exhibits a power-law scaling with the channel height H, 
as shown by the log–log plot in figure~\ref{fig:Fit}(III).

Figure~\ref{fig:Fit}(I) compares our analytical prediction from~\eqref{eq:triple2}
with the MD simulation data of~\citet{qian2006variational},
demonstrating excellent agreement.
This result underscores the importance of the viscous term
$\nabla_{\!\scriptstyle \text{s}}\cdot(\eta\nabla_{\!\scriptstyle \text{s}}\text{u}_{\scriptstyle \text{s}})\delta l$,
which induces an inhomogeneous slip velocity along the fluid–solid interface.
Our formulation therefore complements the interface-formation model of~\cite{shikhmurzaev1993moving},
where slip heterogeneity is attributed solely to variations in surface composition.
In contrast, our model—and the MD results 
of~\citet{qian2003molecular}—show that
even in the absence of compositional gradients, viscous effects 
within the AL can produce comparable slip variations.
Finally, the far-field slip velocity is given by
\begin{align}
\text{U}_{\scriptstyle\infty}=\text{U}_{\scriptscriptstyle 0\,}
\frac{2_{\scriptscriptstyle \,}\delta l}
{\ \text{H}+2_{\scriptscriptstyle \,}\delta l\,},
\label{eq:triple3}
\end{align}
which matches the single-phase Couette flow prediction in~\eqref{eq:u_couette},
as shown in figure~\ref{fig:Fit}(II).
This consistency confirms that, in the far field,
the same interfacial mechanisms governing single-phase flow
remain valid near the moving contact line 
when the adsorption layer is properly accounted for.

In this analysis, we assume equal viscosities for both fluid phases,
consistent with~\cite{qian2006variational}.
If the viscosities differ, however,
the term $\nabla_{\!\scriptstyle\text{s}}\cdot(\eta\nabla_{\!\scriptstyle\text{s}}\mathbf{u}_{\scriptstyle \text{s}})\delta l$
becomes even more significant in the interfacial force balance,
leading to pronounced variations in slippage and slip velocity.
Such cases cannot be solved analytically and will require numerical investigation,
which will be pursued in our future work.

\section{Conclusion}
\textcolor{black}{In this work, we investigate the fundamental
physical mechanisms underlying slip boundary condition
by examining the coupled hydrodynamic and thermodynamic behaviors 
of multicomponent fluids in contact with solid substrates.
We introduced the concept of an adsorption layer (AL),
treating the fluid–solid interface not as a mathematically sharp boundary
but as a finite interfacial region where molecular interactions govern
both mass transport and momentum exchange.
Within this layer, solid–fluid interactions not only drive 
adsorption/depletion—minimizing chemical free energy,
but also induce interfacial flows that dissipate kinetic energy.}

\textcolor{black}{By applying energy minimization principles, 
we established a general force-balance framework within the AL that couples thermodynamic and hydrodynamic effects.
A central outcome is the identification of solid–fluid friction as the key factor controlling slippage.
At the molecular level, this friction originates from the elongation, 
contraction, and rotation of interfacial bonds 
that resist tangential motion, linking the macroscopic frictional force to 
the microscopic bonding energy between solid and fluid molecules.
The AL thus acts as a transitional zone where energy dissipation occurs 
through both hydrodynamic via momentum transfer 
and thermodynamic surface diffusion and convection.}

\textcolor{black}{We validated this framework in classical flow configurations—including 
Poiseuille, Couette, and wedge flows—and demonstrated 
that slip arises as a self-consistent interfacial property.
It depends not only on the interplay between interfacial friction and fluid viscosity in the tangential direction, 
but also on thermodynamic coupling across the interface normal, 
which is typically neglected in sharp-interface models.
This coupling naturally explains phenomena 
such as confinement-enhanced water slippage in carbon nanotubes, 
which cannot be captured by conventional approaches.
In this picture, the “slip length” is no longer a prescribed parameter,
but an emergent, spatially varying quantity determined 
by local friction, viscous drag, and surface composition.}

\textcolor{black}{Together, these results establish a unified framework 
that links microscopic friction within the adsorption layer 
to macroscopic interfacial motion, 
bridging molecular-scale physics with continuum-scale hydrodynamics.
Looking forward, several directions are promising.
Incorporating temperature gradients could capture 
thermally driven slip and Marangoni effects, 
while extending the theory to electrokinetic systems 
may reveal charge-mediated wetting dynamics.}


\section*{Fundings} This work was supported by the Gottfried-Wilhelm Leibniz prize NE 822/31-1 
of the German research foundation (DFG), 
VirtMat project P09 ``Wetting Phenomena'' 
of the Helmholtz Association (MSE-programme No. 43.31.01). 

\section*{Declaration of interests}
The authors report no conflict of interest.

\appendix

\section{Total energy functional and governing equation}
\label{sec:deduction}
Considering the multi-component system with droplets that contact the solid substrate, 
the \emph{fluid} has the total energy density functional $\mathcal{L}$ 
consisting of three parts, namely, 
the chemical free energy functional $\mathcal{F}$, the potential energy $\mathcal{P}$,
and the kinetic energy $\mathcal{K}$,
\begin{gather*}
\mathcal{L} = \mathcal{F} + \mathcal{P} + \mathcal{K}.
\end{gather*}
In addition, the kinetic energy of the \emph{solid substrate} $\mathcal{K}_{\scriptstyle\text{w}}$ 
should also be considered.
Thus, the total energy density functional of the fluid and substrate system reads
\begin{gather*}
\mathcal{L}_{\scriptstyle T} =\mathcal{L} + \mathcal{K}_{\scriptstyle\text{w}}.
\end{gather*}

\subsection{Chemical free energy functional}
The fluid chemical free energy functional $\mathcal{F}$ 
is formulated as the summation of 
the bulk free energy functional ($\mathcal{F}_{\scriptstyle\text{b}}$)
in the bulk fluid domain $\Omega$, 
and the wall energy functional  ($\mathcal{F}_{\scriptstyle\text{s}}$) at the fluid-solid contact surface $\text{S}$ as
\begin{align*}
    \mathcal{F}=\mathcal{F}_{\scriptstyle\text{b}}+\mathcal{F}_{\scriptstyle\text{s}} &= \int_{\scriptstyle\Omega}  
     g
    \,\mathrm{d}\Omega 
    + \int_{\scriptstyle \text{S}} 
    g_{\scriptscriptstyle \text{w}\,}\delta l\,\mathrm{dS}.
\end{align*}
Especially, in this work, we specify the thin layer in the fluid-solid contact region
with thickness $\delta l$
and denominate it as the adsorption layer (AL).
This thin layer denotes the region where solid and fluid interactions exist,
resulting in the adsorption and depletion of the fluid molecules.
For different fluid-solid pairs, the magnitude of $\delta l$ ranges from several nanometer to micro-meter scales~\citep{butt2018adaptive,wang2024wettingb},
which depends not only on the properties of the fluid solid interactions,
but also on the size of the molecules.

\subsubsection{Chemical Free Energy in Bulk Fluids}
The bulk free energy density functional $g(\boldsymbol{c},\nabla\boldsymbol{c})$ takes the Cahn-Hilliard formulation~\citep{cahn1958free,zhang2022janus}
\begin{align*}
    g=f(\boldsymbol{c})
 +{\textstyle\sum}_{\scriptstyle i\,}\kappa\big(\nabla c_{\scriptstyle i}\big)^{\scriptstyle 2}/2,
\end{align*}
in which the spatial-temporal fluid composition is defined 
by $\boldsymbol{c}(\mathbf{x}, \text{t})=( c_{\scriptstyle 1}, c_{\scriptstyle 2},\cdots, c_{\scriptstyle i},\cdots)$
and $f$ is the bulk free energy density which is a function of $\boldsymbol{c}(\mathbf{x}, \text{t})$. 

\subsubsection{Chemical free energy in AL}\label{sec:wall_energy}
In $\mathcal{F}$, 
the second integrand $g_{\scriptstyle \text{w}}(\boldsymbol{c}_{\scriptstyle \text{s}},\nabla\boldsymbol{c}_{\scriptstyle \text{s}})$ 
scales the chemical interaction 
between the fluid and the solid substrate,
and is related with the combination of 
entropy density $s_{\scriptstyle \text{w}}$, 
enthalpy density $h_{\scriptstyle \text{w}}$ 
and van der Waals interaction $f_{\scriptstyle \text{vdW}}$~\citep{wang2024wetting,de1985wetting} as,
\begin{align*}
   f_{\scriptstyle \text{w}} =  h_{\scriptstyle \text{w}}(\boldsymbol{c}) -\,T s_{\scriptstyle \text{w}}(\boldsymbol{c})
    +f_{\scriptstyle \text{vdW}}.
\end{align*}
Noteworthily, $f_{\scriptscriptstyle \text{vdW}}$ between solid and fluid molecules 
alters the free energy of fluid molecules 
and can be calculated with the Lennard-Jones type of potential energy density as
\begin{align*}
    f_{\scriptstyle \text{vdW}} = 
    {\textstyle\sum}_{\scriptstyle i\,}\epsilon_{\scriptscriptstyle\,}c_{\scriptstyle i\,}\big[(\sigma_{\scriptstyle i}/\text{r}_{\scriptstyle 0})^{\,\scriptstyle m}-(\sigma_{\scriptstyle i}/\text{r}_{\scriptstyle 0})^{\,\scriptstyle n}\big],
\end{align*}
where $\epsilon$ is the depth of the potential well
and $\sigma_{\scriptscriptstyle\text{i}}$ is the distance at which the potential energy is zero.
The scaling parameters $m$ and $n$ are obtained from experiments.
Apparently, we assume that the total van der Waals force is composition dependent.
In this way, the wall energy density also depends on the surface composition, namely $f_{\scriptstyle \text{w}}(\boldsymbol{c}_{\scriptstyle\text{s}})$
which differs from the bulk free energy density $f(\boldsymbol{c})$.
By defining the AL thickness $\delta l$ as the characteristic length scale
of the solid-fluid intermolecular forces, 
the total chemical free energy functional
is formulated with
\begin{align*}
    \mathcal{F}
    &= \int_{\scriptstyle \Omega} 
    \Big[f(\boldsymbol{c})
+{\textstyle\sum}_{\scriptstyle i}\kappa\big(\nabla c_{\scriptstyle i}\big)^{\scriptstyle 2}/2\Big]
    \,\mathrm{d}\Omega 
    +\!\! \int_{\scriptstyle \text{S}} 
    \Big[f_{\scriptstyle \text{w}}(\boldsymbol{c}_{\scriptstyle\text{s}})
    +{\textstyle\sum}_{\scriptstyle i}\kappa_{\scriptstyle\text{s}}\big(\nabla c_{\scriptstyle\text{s},i}\big)^{\scriptstyle 2}/2\Big]\delta l
    \,\mathrm{dS}.
\end{align*}

\subsection{Potential energy and kinetic energy}
For the multi-component fluid flow, 
the potential energy of the fluid is represented by 
\begin{align*}
 \mathcal{P}=\mathcal{P}_{\scriptstyle\text{b}}+\mathcal{P}_{\scriptstyle\text{s}} = \int_{\scriptstyle \Omega} p \,\mathrm{d}\Omega +
 \int_{\scriptstyle \text{S}} p_{\scriptstyle\text{s}} \,\delta l\, \mathrm{dS}.
\end{align*}
Here, the pressure density in bulk and AL
are labeled by $p$ and $p_{\scriptstyle\text{s}}$, respectively.
The fluid kinetic energy takes the following formulation 
with the macroscopic fluid velocity $\mathbf u$ inside the bulk region,
and $\mathbf{u}_{\scriptstyle\text{s}}$ in AL,
\begin{align*}
 \mathcal{K}=\mathcal{K}_{\scriptstyle\text{b}}+\mathcal{K}_{\scriptstyle\text{s}} = \int_{\scriptstyle\Omega} \frac{\rho\,\mathbf u^{\scriptstyle 2}}{2} \,\mathrm{d}\Omega +
 \int_{\scriptstyle\text{S}} \frac{\rho \,\mathbf{u}_{\scriptstyle\text{s}}^{\scriptstyle 2}}{2}\delta l\, \mathrm{dS},
\end{align*}
where the density $\rho$ is assumed to be constant.

\subsection{Evolution equations by energy minimization}\label{sec:energy_law}

\subsubsection{Chemical Free Energy dissipation}
From the temporal decay of the chemical free energy functional $\mathcal{F}$, 
we have
\begin{align}
\!\!\!\mathrm{d}\mathcal F/\mathrm{d}\text{t}
 =&\!\! \int_{\scriptstyle \Omega} 
 {\textstyle\sum}_{\scriptstyle i}\Big[\frac{\delta g}{\partial c_{\scriptstyle i}}\frac{\mathrm{d}c_{\scriptstyle i}}{\mathrm{d}\text{t}} 
+\nabla\! \cdot\!\big(\kappa \nabla c_{\scriptstyle i} \otimes \nabla  c_{\scriptstyle i}\big)\cdot\! \mathbf u \Big]\mathrm{d}\Omega
\label{eq:3.1}\\
+&\!\!\int_{\scriptstyle \text{S}} 
{\textstyle\sum}_{\scriptstyle i}\Big[\frac{\delta g_{\scriptstyle \text{w}}}{\partial c_{\scriptstyle\text{s},i}}\frac{\mathrm{d} c_{\scriptstyle\text{s},i}}{\mathrm{d}\text{t}\,}
+\nabla\cdot\big(\kappa_{\scriptstyle \text{s}} \nabla c_{\scriptstyle\text{s},i} \otimes \nabla c_{\scriptstyle\text{s},i}\big)\cdot\mathbf{u}_{\scriptstyle \text{s}} \Big] \delta l_{\scriptscriptstyle\,} \mathrm{dS}.
\label{eq:df_dt}
\end{align}
The second integrand in  \eqref{eq:3.1} is obtained from the divergence theorem and 
the equality of vector calculus $\nabla \cdot (\textbf{a}\otimes\textbf{b})\!=\! \nabla\cdot (\textbf{a})_{\scriptscriptstyle\,}\textbf{b}+ \textbf{a}\cdot \nabla \textbf{b}$.
Physically, Eqs.~\eqref{eq:3.1} and~\eqref{eq:df_dt} 
describe the dissipation of chemical free energy  
via bulk diffusion 
and surface diffusion in AL, respectively.

\subsection{Kinetic energy dissipation}
The kinetic energy dissipation is calculated with the total derivatives as,
\begin{align}
\mathrm{d}\mathcal K/\mathrm{d}\text{t}
= &\int_{\scriptstyle \Omega} \rho_{\scriptscriptstyle \,}\mathbf{u}\cdot
\mathrm{d}\mathbf{u}/\mathrm{d}\text{t} \, \mathrm{d}\Omega
 +\int_{\scriptstyle \text{S}} \rho_{\scriptscriptstyle \,}\mathbf{u}_{\scriptstyle \text{s}}\cdot
 \mathrm{d}\mathbf{u}_{\scriptstyle \text{s}}/\mathrm{d}\text{t}\,\delta l\,\mathrm{dS}.\label{eq:dk_dt0}
\end{align}

\subsection{Potential energy dissipation}
The potential energy dissipation follows the simple chain rule,
\begin{align}
 \mathrm{d}\mathcal{P}/\mathrm{d}\text{t} = \int_{\scriptstyle \Omega} \frac{\mathrm{d} p}{\mathrm{d} \mathbf{x}}\cdot \frac{\mathrm{d} \mathbf{x}}{\mathrm{dt}} \,\mathrm{d}\Omega +
 \int_{\scriptstyle \text{S}} \frac{\mathrm{d}p_{\scriptstyle s}}{\mathrm{d}\mathbf{x}}\cdot \frac{\mathrm{d} \mathbf{x}}{\mathrm{dt}} \,\delta l\, \mathrm{dS}
 =\int_{\scriptstyle \Omega} \nabla p\cdot \mathbf{u} \,\mathrm{d}\Omega +
 \int_{\scriptstyle \text{S}} \nabla p_{\scriptstyle \text{s}}\cdot \mathbf{u}_{\scriptstyle \text{s}}\, \delta l\,\mathrm{dS}.
\label{eq:dp_dt}
\end{align}

\subsection{Governing equations in bulk fluids}\label{sec:GE_bulk}
Collecting the bulk total energy dissipation terms
in~\eqref{eq:df_dt}~\eqref{eq:dk_dt0}~\eqref{eq:dp_dt}, we have
\begin{gather}
\frac{\mathrm{d}\mathcal L_{\scriptstyle \text{b}}}{\mathrm{d}\text{t}} 
 = \frac{\mathrm{d}(\mathcal F_{\scriptstyle \text{b}}+\mathcal K_{\scriptstyle \text{b}}+\mathcal P_{\scriptstyle \text{b}})}{\mathrm{d}\text{t}} 
 \!=\!\! \int_{\scriptstyle \Omega} 
\bigg\{{\textstyle\sum}_{\scriptstyle i}\frac{\delta g}{\delta c_{\scriptstyle i}}
  \frac{\mathrm{d} c_{\scriptstyle i}}{\mathrm{d}\text{t}}
+\Big[{\textstyle\sum}_{\scriptstyle i}\nabla\! \cdot\!\big(\kappa \nabla c_{\scriptstyle i}\! \otimes\! \nabla  c_{\scriptstyle i}\big)
\!+\!\nabla p
\!+\!\rho\frac{\mathrm{d} \mathbf{u} }{\mathrm{d}\text{t}}\Big]\!\cdot\! \mathbf{u}
\bigg\} \mathrm{d}\Omega. \label{eq:dLb_dt}
\end{gather}
We easily obtain the evolution equations by the energy dissipation in bulk fluid---
the Cahn-Hilliard-Navier-Stokes model (CHNS), or the so-called model H,
\begin{equation}
\begin{gathered}
\nabla \cdot\mathbf{u}=0,\\[0.3em]
\partial c_{\scriptstyle i}/\partial\text{t}
+\mathbf{u}\cdot\nabla c_{\scriptstyle i}
=\nabla\cdot\big(M\nabla \mu_{\scriptstyle i}\big),\\[0.4em]
\rho_{\scriptstyle \,}\mathrm{d}\mathbf u/\mathrm{d}\text{t} 
= -\nabla \cdot \big(\,\text{P}\textbf{I} + \underline{\underline{\Theta}}\,\big)
+\nabla \cdot\underline{\underline{\boldsymbol{\sigma}}}.
\end{gathered}  \label{eq:GE_bulk}
\end{equation} 
The first equation denotes the incompressibility.
The second evolves the composition in bulk
in which the bulk chemical potential 
is defined with $\mu_{\scriptstyle i} =\delta g/\delta c_{\scriptstyle i}=
\partial f/\partial c_{\scriptstyle i}
 -\kappa\nabla^{\scriptstyle 2}c_{\scriptstyle i}$.
The mobility $M$ scales the bulk diffusion
which follows the Onsager's relation; more details in our previous work~\cite{zhang2021phase}.
The third equation describes the force balance
on the fluid takes the tensor formation,
which contains the grand pressure $\text{P}$,
the surface stress tensor $\underline{\underline{\Theta}}$,
and viscous stress tensor $\underline{\underline{\boldsymbol{\sigma}}}$,
\begin{gather*}
\text{P}= p + g - {\textstyle\sum}_{\scriptstyle i}\frac{\delta g}{\delta c_{\scriptstyle i}}c_{\scriptstyle i},\quad
 \underline{\underline{\Theta}}
 = -\Big(g - {\textstyle\sum}_{\scriptstyle i}\frac{\delta g}{\delta c_{\scriptstyle i}}
 c_{\scriptstyle i}\Big)\mathbf{I} 
 + {\textstyle\sum}_{\scriptstyle i\,}\kappa\,\nabla c_{\scriptstyle i}\otimes\nabla c_{\scriptstyle i}, \quad    \underline{\underline{\boldsymbol{\sigma}}}
 =\eta\,\big(\nabla\mathbf{u} + \nabla\mathbf{u}^{\scriptscriptstyle\text{T}}\big).
\end{gather*}

If not written in the tensor form, 
the force balance equation can also be expressed as
\begin{equation*}
\rho_{\scriptstyle \,}\mathrm{d}\mathbf u/\mathrm{d}\text{t} 
= -\nabla \text{P} - {\textstyle\sum}_{\scriptstyle i}c_{\scriptstyle i}\nabla\mu_{\scriptstyle i}
+\nabla\cdot\big[\eta\,\big(\nabla\mathbf{u} + \nabla\mathbf{u}^{\scriptscriptstyle\text{T}}\big)\big].
\end{equation*} 

Substituting~\eqref{eq:GE_bulk} into~\eqref{eq:dLb_dt}, the total energy dissipation in bulk fluid obeys,
\begin{gather}
\mathrm{d}\mathcal L_{\scriptstyle \text{b}}/\mathrm{d}\text{t}
 =-\int_{\scriptstyle \Omega} 
\Big[{\textstyle\sum}_{\scriptstyle i}M (\nabla \mu_{\scriptstyle i})^{\scriptstyle 2}
+\eta\nabla \mathbf{u} :\nabla \mathbf{u}
\Big]\, \mathrm{d}\Omega. \label{eq:dLb_dt1}
\end{gather}

\subsection{Governing equations in adsorption layer}\label{sec:GE_AL}
Next, we focus on deducing the feasible force balance equation in AL.
Still, the skeleton key is the total energy minimization.
\begin{gather}
 \frac{\mathrm{d}\mathcal L_{\scriptstyle \text{s}}}{\mathrm{d}\text{t}} 
 = \int_{\scriptstyle \text{S}} 
\bigg\{
{\textstyle\sum}_{\scriptstyle i}\frac{\delta g_{\scriptstyle \text{w}}}{\delta c_{\scriptstyle\text{s},i}}
\frac{\mathrm{d} c_{\scriptstyle\text{s},i}}{\mathrm{d}\text{t}\,}
+\Big[{\textstyle\sum}_{\scriptstyle i}\nabla\!\cdot\!\big(\kappa_{\scriptstyle \text{s}} \nabla c_{\scriptstyle\text{s},i} \otimes \nabla c_{\scriptstyle \text{s},i}\big)
+\nabla p_{\scriptstyle\text{s}}
+\rho_{\scriptscriptstyle \,}\frac{\mathrm{d} \mathbf{u}_{\scriptstyle\text{s}} }{\mathrm{d}\text{t}} \Big]
 \cdot \mathbf{u}_{\scriptstyle\text{s}} \bigg\} \delta l_{\scriptscriptstyle\,} \mathrm{dS}.\label{eq:dLs_dt}
\end{gather}
The derivation is similar to the bulk fluid, 
but with only one difference.
The friction force between fluid and solid substrate needs to be considered in the force balance equation in AL,
\begin{equation}
\begin{gathered}
\nabla \cdot\mathbf{u}_{\scriptstyle \text{s}}=0,\\[0.3em]
\partial c_{\scriptstyle\text{s},i}/\partial\text{t}
+\mathbf{u}_{\scriptstyle \text{s}}\cdot\nabla c_{\scriptstyle\text{s},i}
=\nabla\cdot\big(M_{\scriptstyle \text{s}}\nabla \mu_{\scriptstyle\text{s},i}\big),\\[0.4em]
\rho_{\scriptstyle \,}\mathrm{d}\mathbf u_{\scriptstyle \text{s}}/\mathrm{d}\text{t} 
= -\nabla \cdot \big(\,\text{P}_{\scriptstyle \text{s}\,}\textbf{I} 
+ \underline{\underline{\Theta_{\scriptstyle \text{s}}}}\,\big)
+\nabla \cdot\underline{\underline{\boldsymbol{\sigma}_{\scriptstyle \text{s}}}}
-\lambda (\mathbf{u}_{\scriptstyle \text{s}}-\mathbf{u}_{\scriptstyle \text{w}}),
\end{gathered}  \label{eq:GE_AL}
\end{equation} 
where the surface chemical potential 
is defined with $\mu_{\scriptstyle\text{s},i} =\delta g_{\scriptstyle\text{w}}/\delta c_{\scriptstyle\text{s},i}=
\partial f_{\scriptstyle\text{w}}/\partial c_{\scriptstyle\text{s},i}
 -\kappa\nabla^{\scriptstyle 2}c_{\scriptstyle\text{s},i}$.
The mobility $M_{\scriptstyle \text{s}}$ scales the surface diffusion.
The stress tensors are defined with the same formulations as the bulk fluids,
\begin{gather*}
\text{P}_{\scriptstyle\text{s}}= p_{\scriptstyle\text{s}} + g_{\scriptstyle\text{w}} - {\textstyle\sum}_{\scriptstyle i}(\delta g_{\scriptstyle\text{w}}/\delta c_{\scriptstyle\text{s},i})c_{\scriptstyle\text{s},i},\quad\quad\quad\quad
\underline{\underline{\boldsymbol{\sigma}_{\scriptstyle\text{s}}}}
 =\eta\,\big(\nabla\mathbf{u}_{\scriptstyle\text{s}} + \nabla\mathbf{u}_{\scriptstyle\text{s}}^{\scriptscriptstyle\text{T}}\big),\\[0.5em]
 \underline{\underline{\Theta_{\scriptstyle\text{s}}}}
 = -\big[g_{\scriptstyle\text{w}} - {\textstyle\sum}_{\scriptstyle i}(\delta g_{\scriptstyle\text{w}}/\delta c_{\scriptstyle\text{s},i})
 c_{\scriptstyle\text{s},i}\big]\mathbf{I} 
 + {\textstyle\sum}_{\scriptstyle i\,}\kappa\,\nabla c_{\scriptstyle\text{s},i}\otimes\nabla c_{\scriptstyle\text{s},i}.
\end{gather*}
If not written in the tensor form, 
the force balance in AL is expressed as
\begin{equation*}
\rho_{\scriptstyle \,}\frac{\mathrm{d}\mathbf u_{\scriptstyle \text{s}}}{\mathrm{dt}}
= -\nabla p_{\scriptstyle \text{s}}
-\sum_{\scriptstyle i}c_{\scriptstyle \text{s},i}\nabla\mu_{\scriptstyle \text{s},i}
+\nabla\cdot\Big[\eta(\nabla\textbf{u}_{\scriptstyle \text{s}}+\nabla\textbf{u}_{\scriptstyle \text{s}}^{\scriptstyle \text{T}})\Big]
-\lambda (\mathbf{u}_{\scriptstyle \text{s}}\!-\!\mathbf{u}_{\scriptstyle \text{w}}),
\end{equation*}
which recovers the expression we applied in this work.

\subsection{Energy dissipation}\label{sec:dldt}
Substituting~\eqref{eq:GE_AL} into~\eqref{eq:dLs_dt}, the total surface energy dissipation in AL obeys,
\begin{gather}
\frac{\mathrm{d}\mathcal L_{\scriptstyle \text{s}}}{\mathrm{d}\text{t}} 
 =-\int_{\scriptstyle \Omega} 
\Big[{\textstyle\sum}_{\scriptstyle i}M_{\scriptstyle \text{s}} (\nabla \mu_{\scriptstyle\text{s},i})^{\scriptstyle 2}
+\eta\nabla \mathbf{u}_{\scriptstyle \text{s}} :\nabla \mathbf{u}_{\scriptstyle \text{s}}
+\lambda (\mathbf{u}_{\scriptstyle \text{s}}-\mathbf{u}_{\scriptstyle \text{w}})\cdot \mathbf{u}_{\scriptstyle \text{s}}
\Big]\delta l \mathrm{dS}. \label{eq:dLb_dt2}
\end{gather}
This term is not always negative, because the fluid-solid friction can do work on fluid
and can either accelerate, or decelerate the fluid.
So we shall not forget the power made by the solid-fluid friction onto the \emph{solid substrate}, namely
\begin{align*}
\frac{\mathrm{d} \mathcal K_{\scriptstyle \text{w}}}{\mathrm{dt}}
=& \!\int_{\scriptstyle \text{S}} 
 \!-_{\scriptscriptstyle \,}\textbf{F}\cdot \mathbf u_{\scriptstyle\, \text{w}}\,\delta l\,\mathrm{dS}
= \!\int_{\scriptstyle \text{S}} 
 \lambda \,(\mathbf{u}_{\scriptstyle\text{s}}-\mathbf{u}_{\scriptstyle\text{w}}) \cdot \mathbf u_{\scriptstyle \text{w}}\,\delta l\,\mathrm{dS}.
\end{align*}
Combining $\mathrm{d} \mathcal L/\mathrm{d}\text{t}$ with
$\mathrm{d} \mathcal K_{\scriptstyle \text{w}}/\mathrm{d}\text{t}$,
the total energy functional of the 
\emph{multicomponent fluid and solid substrate} dissipates with time as,
\begin{align*}
\frac{\mathrm{d} \mathcal L_{\scriptstyle \text{T}}}{\mathrm{d}\text{t}}
= \frac{\mathrm{d} (\mathcal L_{\scriptstyle \text{b}}+\mathcal L_{\scriptstyle \text{s}}+\mathcal K_{\scriptstyle \text{w}})}{\mathrm{d}\text{t}} 
=& -\!\!\int_{\scriptstyle \Omega} 
\big[{\textstyle\sum}_{\scriptstyle i}M \big(\nabla \mu_{\scriptstyle i} \big)^{\scriptstyle 2} +  \eta\nabla\mathbf u :\!\nabla\mathbf u\big]\,\mathrm{d}\Omega \\
&-\! \!\int_{\scriptstyle \text{S}} 
 \big[{\textstyle\sum}_{\scriptstyle i}M_{\scriptstyle\text{s}} (\nabla\mu_{\scriptstyle\text{s},i})^{\scriptstyle\, 2}\! +  \eta
 \nabla\mathbf{u}_{\scriptstyle\text{s}}  :\!\nabla\mathbf{u}_{\scriptstyle\text{s}} 
 +\lambda (\mathbf{u}_{\scriptstyle\text{s}}-\mathbf{u}_{\scriptstyle\text{w}})^{\scriptstyle 2}\big]\delta l\,\mathrm{dS}
\leq 0  \quad\blacksquare
\end{align*}
The first integral in domain $\Omega$ denotes the energy dissipation in bulk fluids 
via diffusion and fluid viscous effect.
The second integral in AL highlights the three feasible energy minimizing mechanisms,
namely, the surface diffusion (depletion/adsorption), the fluid viscous effect, and the solid-fluid friction.

\section{Composition Evolution Equation in Adsorption Layer} 
Inside AL, we use the Cahn-Hilliard type
the composition conservation equation,
\begin{equation}
\frac{\,\mathrm{d} c_{\scriptstyle \text{s}}}{\mathrm{dt}} 
=\nabla\cdot(M_{\scriptstyle\text{s}}\nabla\mu_{\scriptstyle\text{s}}),
 \label{eq:GE_AL1_appendix}
\end{equation} 
which is different from the non-conserved form used by~\cite{shikhmurzaev1993moving,jacqmin2000contact,qian2006variational,yue2010sharp} as
\begin{equation}
\frac{\,\mathrm{d} \tilde{c}_{\scriptstyle \text{s}}}{\mathrm{dt}} 
=-\tau\tilde{\mu}_{\scriptstyle\text{s}},\label{eq:wbc_con}
\end{equation} 
where the surface 
chemical potential in the sharp interface model $\tilde{\mu}_{\scriptstyle\text{s}}$
is introduced as a pure mathematic concept 
with the unit of $\si{J/m^{\scriptstyle 2}}$.
While the chemical potential in the present AL approach now has a clear physical meaning
which is related to the interacting solid and fluid molecules
and has the unit of $\si{J/m^{\scriptstyle 3}}$.
The composition in the sharp interface model $\tilde{c}_{\scriptstyle\text{s}}$
has the relation with our clearly defined surface composition
$c_{\scriptstyle\text{s}}$ 
as $\tilde{c}_{\scriptstyle\text{s}}=c_{\scriptstyle\text{s}}\delta l$.

\subsection{Wetting boundary condition revisit}\label{sec:wbc_revisit}
In the following parts, we proof that the conventional non-conserved 
wetting boundary condition~\eqref{eq:wbc_con}
is only a numerical approach
which is a simplification of our model.
\begin{figure}
    \centering
    \includegraphics[width=0.9\textwidth]{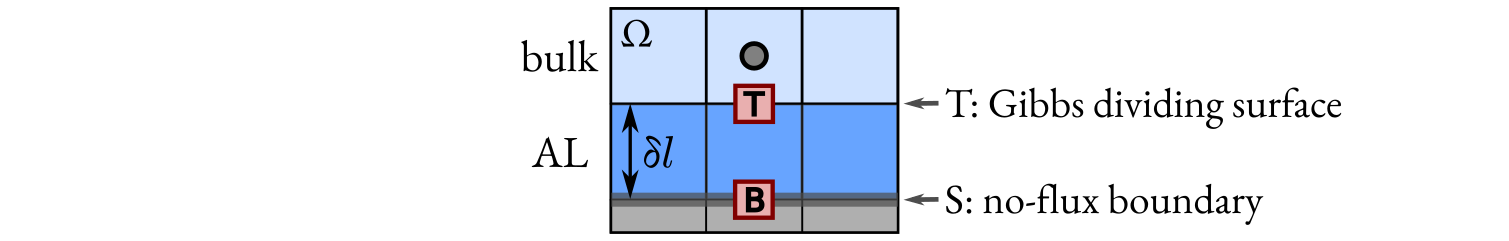} 
    \caption{Discretization of the bulk domain $\Omega$ and adsorption layer (AL).}
    \label{fig:grid}
\end{figure}
First, we discretize the diffusion flux in adsorption layer
into the substrate normal (n) 
and tangential (t) directions, respectively,
\begin{align}
\frac{\,\mathrm{d} c_{\scriptstyle \text{s}}}{\mathrm{dt}} 
&=\nabla_{\!\scriptstyle \text{n}}\!\cdot\!\big(M_{\scriptstyle\text{s}}\nabla_{\!\scriptstyle \text{n}}\,\mu_{\scriptstyle\text{s}}\big)
+\nabla_{\!\scriptstyle\text{t}}\!\cdot\!\big(M_{\scriptstyle\text{s}}\nabla_{\!\scriptstyle\text{t}}\,\mu_{\scriptstyle\text{s}}\big).
\end{align} 

\emph{Ansatz 1:} If we neglect the diffusion flux in the substrate tangential direction,
and let the surface mobility $M_{\scriptstyle\text{s}}$ 
to be a constant
independent on the surface composition,
the material conservation equation is simplified as
\begin{align*}
\frac{\,\mathrm{d} c_{\scriptstyle \text{s}}}{\mathrm{dt}} 
&=\nabla_{\!\scriptstyle \text{n}}\!\cdot\!\big(M_{\scriptstyle\text{s}}\nabla_{\!\scriptstyle \text{n}}\,\mu_{\scriptstyle\text{s}}\big)
=\frac{M_{\scriptstyle\text{s}}\nabla_{\!\scriptstyle \text{n}}\,\mu_{\scriptstyle\text{s}}^{\scriptstyle\text{T}} 
- M_{\scriptstyle\text{s}}\nabla_{\!\scriptstyle \text{n}}\,\mu_{\scriptstyle\text{s}}^{\scriptstyle\text{B}}}{\delta l},
\end{align*}
in which the fluxes at top, bottom are positioned at
the colored squares in figure~\ref{fig:grid}
and represented by T, B, respectively.
Noting the no flux boundary condition 
at the real solid-fluid interface (position B),
\begin{align*}
\frac{\,\mathrm{d} c_{\scriptstyle \text{s}}}{\mathrm{d}\text{t}} 
&=\frac{M_{\scriptstyle\text{s}}}{\delta l}\nabla_{\!\scriptstyle \text{n}}\,\mu_{\scriptstyle\text{s}}^{\scriptstyle\text{T}} 
=-\frac{\,M_{\scriptstyle\text{s}}\,\,}{\delta l^{\scriptstyle 2}}
\big(\mu_{\scriptstyle\text{s}} -\mu^{\scriptstyle \text{o}}\big),
\end{align*}
where $\mu^{\scriptstyle \text{o}}$ is the chemical potential of the cell above AL, depicted by the black dot in figure~\ref{fig:grid}.

\emph{Ansatz 2:} Neglecting the thermodynamical coupling between
AL and the bulk fluid by chemical potential
with $\mu^{\scriptstyle \text{o}}=0$, we have
\begin{align*}
\frac{\,\mathrm{d} c_{\scriptstyle \text{s}}}{\mathrm{dt}} 
=-\frac{\,M_{\scriptstyle \text{s}}\,\,}{\delta l^{\scriptstyle 2}}\Big[\frac{\kappa_{\scriptstyle \text{s}}\nabla_{\!\scriptstyle \text{n}} c_{\scriptstyle \text{s}}^{\scriptstyle\text{T}}
}{\delta l}
+\nabla_{\!\scriptstyle \text{t}}\cdot(\kappa_{\scriptstyle \text{s}}\nabla_{\!\scriptstyle \text{t}\,}c_{\scriptstyle \text{s}})
-\frac{\partial f_{\scriptstyle \text{w}}}{\partial c_{\scriptstyle \text{s}}}\Big].
\end{align*} 

\emph{Ansatz 3:} Neglecting the surface tangential diffusion  
with $\nabla_{\!\scriptstyle \text{t}}\cdot(\kappa_{\scriptstyle \text{s}}\nabla_{\!\scriptstyle \text{t}\,}c_{\scriptstyle \text{s}})=0$, 
the standard non-conserved Allen-Cahn type of wetting boundary condition~\citep{jacqmin2000contact, ding2007wetting,wang2008moving,carlson2009modeling,yue2010sharp} is recovered
from the conserved Cahn-Hilliard formulation,
\begin{align}
\frac{\,\mathrm{d} c_{\scriptstyle \text{s}}}{\mathrm{dt}} 
&=-\frac{\,M_{\scriptstyle \text{s}}\,\,}{\delta l^{\scriptstyle 3}}\Big(\kappa_{\scriptstyle \text{s}}\nabla_{\!\scriptstyle \text{n}} c_{\scriptstyle \text{s}}^{\scriptstyle\text{T}}
-\delta l\,\frac{\partial f_{\scriptstyle \text{w}}}{\partial c_{\scriptstyle \text{s}}}\Big)
=-\tau \Big(\kappa\nabla c_{\scriptstyle \text{s}}\cdot \mathbf{n}
-\frac{\partial \gamma}{\partial c_{\scriptstyle \text{s}}}\Big).\label{eq:wbc_ac}
\end{align} 
where the term $\gamma(c_{\scriptstyle \text{s}}):=f_{\scriptstyle \text{w}}
(c_{\scriptstyle \text{s}}){\scriptscriptstyle\,}\delta l$ 
is understood as the solid-fluid interfacial tension
which is a function of $c_{\scriptstyle \text{s}}$.
Hence, the so-called phenomenological kinetic parameter 
$\tau$ in previous models has a concrete physical
definition as
\begin{align*}
    \tau = M_{\scriptstyle \text{s}}/\delta l^{\scriptstyle 3},
\end{align*}
which scales the diffusion flux of fluid molecules
entering the adsorption layer 
from the bulk region
through the surface T,
due to the unbalanced chemical potential.
Besides, $\tau$ shows
a clear relationship with the characteristic length scale $\delta l$ 
which denotes the range of solid-fluid chemical
interactions.

It needs to be stress that
Ansatz 2 is not always true for many systems,
even when the equilibrium of AL is reached. 
Actually, the static state indicates the equal chemical potential
\begin{align*}
\mu_{\scriptstyle \text{s}}
-\mu^{\scriptstyle \text{o}}
=\nabla \mu_{\scriptstyle \text{s}} \cdot \mathbf{n}\,\Big\rvert_{\scriptstyle \text{T}}=0,
\end{align*}
which is the real no-flux boundary condition at the position T.
However, the Allen-Cahn type of wetting boundary condition~\eqref{eq:wbc_ac} is a special case 
which only leads to $\mu_{\scriptstyle \text{s}}=\mu^{\scriptstyle \text{o}}=0$ 
at equilibrium states.~\eqref{eq:wbc_ac} works well for the droplet wetting and hydrodynamic scenarios, 
because of its subtle influence compared with the strong hydrodynamics.
However, the constraint $\mu_{\scriptstyle \text{s}}=\mu^{\scriptstyle \text{o}}=0$ 
is not always thermodynamically consistent.

Easily picking a counterexample, 
for the $1:1$ binary electrolyte with initial composition $c_{\scriptstyle 0}$,
the equilibrium electrochemical potential of the ions in HDL 
is identical to $\mu$ at the center of the electrolyte 
where the electric potential $\Psi=0$, yielding
\begin{align*}
  \mu_{\scriptstyle \text{s}}=\mu^{\scriptstyle \text{o}}
  =\frac{k_{\scriptscriptstyle
   B} T}{\delta l^{\scriptstyle 3}}
   \ln\Big(\frac{c_{\scriptscriptstyle 0}}
   {1-2\,c_{\scriptscriptstyle 0}}\Big)\neq 0.
\end{align*}

\subsection{Young's law}\label{sec:sharp}
In this part, we present our wetting boundary model
is thermodynamically consistent with previous sharp interface models.
\begin{figure}
\includegraphics[width=0.9\textwidth]{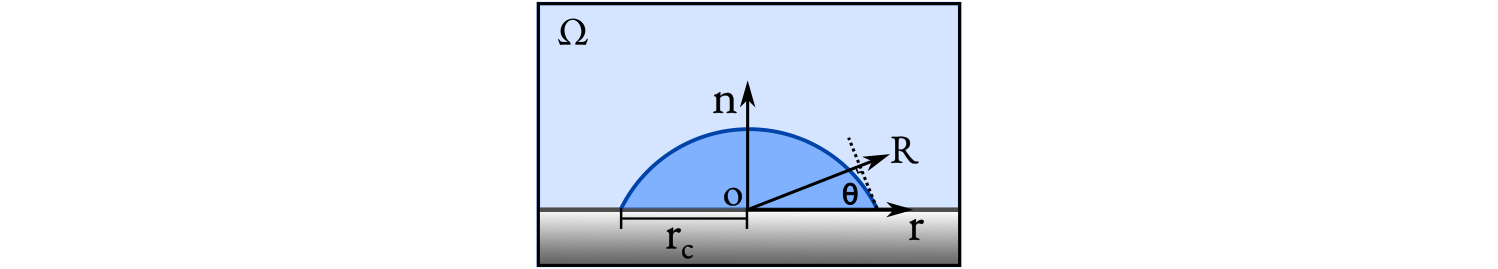}
\centering
\caption{Sessile droplet on the substrate. 
$\text{r}_{\scriptstyle \text{c}}$ is the contact line radius, 
R is the normal direction of the droplet-air surface.
}
\label{fig:domain}
\end{figure}
we derive the Young's law.
As the sessile droplet illustrated in figure~\ref{fig:domain},
integrating $\mu_{\scriptstyle \text{s}}\,\delta l$
in the cylindrical coordinates
from the equilibrium surface liquid composition $c_{\scriptscriptstyle 0}$ 
at the droplet center position $\text{r}=0$
to the equilibrium surface air composition $c_{\scriptscriptstyle \infty}$ at $\text{r}=\infty$,
yielding
\begin{align}
\int_{c_{\scriptscriptstyle 0}}^{c_{\scriptscriptstyle \infty}}\!\!
\mu_{\scriptstyle \text{s}}\,\delta l \,
\mathrm{d}c_{\scriptstyle \text{s}}
 &= f_{\scriptstyle \text{w}}(c_{\scriptstyle \infty})\delta l
 -f_{\scriptstyle \text{w}}(c_{\scriptstyle 0})\delta l
-\int_{\scriptstyle c_{0}}^{\scriptstyle c_{\infty}}
\!\!\bigg[\kappa_{\scriptstyle \text{s}}\frac{\partial c_{\scriptstyle \text{s}}}{\partial \text{r}}
\cos\theta
+\frac{\kappa_{\scriptstyle \text{s}}}{\text{r}}
\frac{\partial}{\partial \text{r}}\Big(\text{r}
\frac{\;\partial c_{\scriptstyle \text{s}}}{\partial \text{r}}\Big)
 \delta l\bigg]\mathrm{d}c_{\scriptstyle \text{s}}\notag\\
 &= \gamma(c_{\scriptstyle \infty})
 -\gamma(c_{\scriptstyle 0})
-\sigma\cos\theta
 -\int_{\scriptstyle 0}^{\scriptstyle \infty}
 \frac{\,\kappa_{\scriptstyle \text{s}}}{\text{r}}
 \Big(\frac{\;\partial c_{\scriptstyle \text{s}}}
 {\partial \text{r}}\Big)^{\scriptstyle 2}
 \delta l\,\mathrm{dr} \notag\\
 &= \Delta\gamma
-\sigma\cos\theta
 -\frac{\,\tau_{\scriptscriptstyle \text{L}}\,\delta l\,}{\text{r}_{\scriptstyle\text{c}}},\label{eq:mu_dr}
\end{align}
where the liquid-air interfacial tension reads 
$\sigma:={\scriptstyle\int}_{\scriptstyle 0}^{\scriptstyle \infty}
\kappa(\partial_{\scriptstyle \text{r}}c_{\scriptstyle \text{s}})^{\scriptstyle 2\,}\mathrm{dr}$,
which is integrated along the droplet-air surface normal direction R; see figure~\ref{fig:domain}.
Meanwhile, the solid-liquid and solid-air interfacial tensions 
are represented by $\gamma(c_{\scriptstyle 0})$, 
$\gamma(c _{\scriptstyle\infty})$, respectively.
Hence, $\Delta \gamma=\gamma(c _{\scriptstyle\infty})-\gamma(c_{\scriptstyle 0})$.
The contact line radius is $\text{r}_{\text{c}}$.
More details on the parameter 
$\tau_{\scriptscriptstyle \text{L}}:={\scriptstyle\int}_{\scriptstyle 0}^{\scriptstyle \infty}
\kappa_{\scriptstyle \text{s}}(\partial_{\scriptstyle \text{r}}
c_{\scriptstyle \text{s}})^{\scriptstyle 2}\mathrm{dr}$ 
is presented in our previous work~\cite{zhang2023line}.

Letting~\eqref{eq:mu_dr} equal to $0$, 
the Young's law with line tension effect~\citep{amirfazli2004status,law2017line} is recovered,
\begin{align}
    \cos\theta = \frac{\Delta \gamma}{\sigma} 
    - \frac{\,\tau_{\scriptscriptstyle \text{L}}\,\delta l\,}{\sigma_{\scriptstyle \,}\text{r}_{\text{c}}}+\mathcal{O}(\text{r}_{\text{c}}^{\scriptstyle \,-2}).\label{eq:young}
\end{align}
Noteworthily,~\eqref{eq:young} indicates the physical meaning 
of AL thickness $\delta l$
from another perspective.
In this work, 
we define the AL as the range of solid-fluid interactions
from the energy perspective.
Hence, at the distance farer than $\delta l$ in the fluid bulk region, 
the fluid molecules are affected by the fluid-fluid interactions only, 
and no solid-fluid interaction exists. 
Consequently, AL and bulk fluid are intrinsically separated by the surface T
locating at the position $\delta l$ above the solid-fluid interface; see figure~\ref{fig:grid}.

Based on this understanding, 
we suggest that the plane T is equivalent to 
the Gibbs dividing surface~\citep{gibbs1928collected,schimmele2007conceptual}.
Therefore, $\delta l$ scales the magnitude of line tension effect
which is in accordance with previous studies~\citep{zhang2023line,wang2024wetting}.

\bibliographystyle{jfm}
\bibliography{jfm}

\end{document}